\documentclass[12pt]{article}
\usepackage{amsmath}
\usepackage{graphicx}
\usepackage{bm}
\usepackage{fancyhdr}
\usepackage{amssymb}
\usepackage{setspace}
\usepackage{color}
\usepackage{epsfig}
\usepackage{overpic}
\usepackage{caption,subcaption}
\usepackage{epstopdf}
\usepackage{ulem}
\begin{document}

\baselineskip=20pt

\newcommand{\Title}[1]{{\baselineskip=26pt
   \begin{center} \Large \bf #1 \\ \ \\ \end{center}}}
\newcommand{\Author}{\begin{center}
   \large \bf
Wei Wang${}^{a,b}$, Yi Qiao${}^{b}$, Junpeng Cao${}^{b,c,d,e}$, Wu-Ming Liu${}^{b,c,d}\footnote{Corresponding author: wmliu@iphy.ac.cn}$ and Rong-Hua Liu${}^{a} \footnote{Corresponding author: rhliu@nju.edu.cn}$
 \end{center}}
\newcommand{\Address}{\begin{center}
     ${}^a$National Laboratory of Solid State Microstructures, School of Physics and Collaborative Innovation Center of Advanced
           Microstructures, Nanjing University, Nanjing 210093, China\\
     ${}^b$Beijing National Laboratory for Condensed Matter Physics, Institute of Physics, Chinese Academy of Sciences, Beijing
           100190, China\\
     ${}^c$School of Physical Sciences, University of Chinese Academy of Sciences, Beijing 100049, China\\
     ${}^d$Songshan Lake Materials Laboratory, Dongguan, Guangdong 523808, China\\
     ${}^e$Peng Huanwu Center for Fundamental Theory, Xian 710127, China\\
   \end{center}}

\Title{Exact ground state and elementary excitations of a competing spin chain with twisted boundary condition} \Author

\Address \vspace{0.1cm}

\vspace{1truecm}

\begin{abstract}
A novel Bethe ansatz scheme is proposed to investigate the exact physical properties
of an integrable anisotropic quantum spin chain with competing interactions among the nearest, next nearest neighbor and
chiral three spin couplings, where the boundary condition is the twisted one.
The eigenvalue of the transfer matrix is characterized by its zero roots instead of the traditional Bethe roots. Based on the exact solution,
the conserved momentum and charge operators of this $U(1)$-symmetry broken system are obtained.
The ground state energy and density of rapidities are calculated. It is found that there exist three kinds of elementary excitations and the corresponding dispersion relations are obtained,
which gives a different picture from that with periodic boundary condition.
\vspace{1truecm}


\end{abstract}

\newpage

\hbadness=10000

\tolerance=10000

\hfuzz=150pt

\vfuzz=150pt
\section{Introduction}
\setcounter{equation}{0}

The study of frustrated spin systems is an important and fascinating issue in the condensed matter physics \cite{Balents}.
Due to the competition of various interactions, many interesting phenomena such as different magnetic ordered states are induced in this kind of system \cite{2}.
A typical one-dimensional frustrated model is the $J_1-J_2$ spin chain where the nearest-neighbor (NN) and the next-nearest-neighbor (NNN) couplings are included \cite{Sha81}.
Besides the spin-exchanging interaction, the chiral three spins coupling is another important interaction.
In 1987, Kalmeyer and Laughlin presented that the spin-singlet ground state of a frustrated Heisenberg antiferromagnet can describe the fractional quantum Hall state \cite{Laughlin87},
which is known as the chiral spin liquid. Later, Wen et al \cite{wen89} and Baskaran \cite{Bask89} clarified that the expectation value of the chiral three-spin operator can be used as the order parameter of spin liquid states. After that, the related physical properties such as spin-charge separation, bulk gap and chiral edge states are studied extensively
\cite{Fradkin91,Yang93}. The quantum liquid behavior of one-dimensional system is a very interesting issue \cite{Imam12}
and the pseudoparticle approach is a powerful method to exactly calculate the physical quantities based on the integrable models \cite{Car18}. Recently, the models with spin chirality terms cause renewed attention in the condensed matter physics \cite{Nom92,93,97,98}, statistical mechanics \cite{Djo16} and quantum field theory \cite{Jaf06,Goro15,Chen17}.

At present, the exact energy spectrum of the one-dimensional $J_1-J_2$ model is still an open question. However,
when adding the spin chirality terms, the generalized $J_1-J_2$ model can be mapped into a two quantum spin chains coupled system and is
exactly solvable \cite{Pop93,Zvy95-1,Zvy95-2}. For example, Frahm et al constructed an integrable family of coupled Heisenberg spin chains \cite{Frahm96} and studied the zero-temperature properties utilizing algebraic Bethe ansatz \cite{Frahm97}. Ikhlef, Jacobsen and Saleur constructed an integrable $Z_2$ staggered vertex model and explained the possible applications \cite{IK08,IK10}.
Other interesting developments in this direction can be found in the references \cite{Arna00,Zvy01}.

The integrable boundary conditions of the  generalized $J_1-J_2$ model are the periodic, antiperiodic and open ones.
When we consider the antiperiodic or the non-diagonal boundary conditions, the $U(1)$ symmetry of the system is broken and the coordinate and algebraic Bethe ansatz methods
do not work because of lacking the reference state.
Due to the extensive applications in open string theory, non-equilibrium statistical mechanics and topological physics,
the exact solutions of quantum integrable systems without $U(1)$ symmetry are very important.
Several methods such as gauge transformation \cite{cao03}, $T-Q$ relation based on the fusion \cite{Yung95,nep021}, $q$-Onsager algebra \cite{Bas1,Bas2},
separation of variables \cite{sk2-2,Niccoli13-1}, modified algebraic Bethe ansatz \cite{Bel13-1,Bel13-2,Bel13-3}
and off-diagonal Bethe ansatz \cite{cysw,Book} have been proposed to overcome this obstacle.
Focus on the generalized $J_1-J_2$ model, the eigenvalue of the transfer matrix is characterized by the inhomogeneous $T-Q$ relation, where the associated Bethe ansatz equations (BAEs)
are also inhomogeneous \cite{Qiao20}.
Then another problem arises, that is it is quite hard to calculate the physical quantities in the thermodynamic limit because of the existence of the inhomogeneous term.
In order to solve this difficulty, the $t-W$ relation is proposed \cite{Qiao2020}. Based on it, the elementary excitations and surface energy of
XXZ spin chain with antiperiodic boundary condition are obtained.

In this paper, we study a more general quantum spin chain that includes the NN, NNN and chiral three-spin interactions.
We consider the antiperiodic boundary condition. We find that if we parameterize the eigenvalue of the transfer matrix by its zero roots,
we could obtain the homogeneous BAEs, which makes it possible for us to take the thermodynamic limit and calculate the physical quantities exactly.
We remark that the fusion relation can give all the necessary information to determine the energy spectrum.
Based on them, we calculate the exact physical quantities in the thermodynamic limit.
We obtain the ground state energy, elementary excitations and corresponding dispersion relations.
The conserved momentum and charge operators of this $U(1)$-symmetry broken system are also given.

The paper is organized as follows. In the next section, we introduce the model Hamiltonian and explain its integrability.
In section 3, we show how to parameterize the eigenvalue of the transfer matrix and how to obtain the
homogeneous BAEs.
In section 4, we derive the conserved momentum and charge operators of the system.
The ground state energy with real $a$ and imaginary $\eta$ is derived and
the thermodynamic limit is calculated in section 5. In section 6, three kinds of elementary excitations and
corresponding excited energies are deduced.
Concluding remarks and discussions are given in section 7.

\section{The system and its integrability}
\setcounter{equation}{0}

We consider an anisotropic quantum spin chain which includes the NN, NNN and
chiral three-spin interactions with the antiperiodic boundary condition. The model Hamiltonian reads
\begin{eqnarray}\label{Ham1}
H =-\sum^{2N}_{j=1} \sum_{\alpha=x,y,z}  \big[ J_1^\alpha \sigma_j^\alpha \sigma_{j+1}^\alpha
+J_2 \sigma_j^\alpha \sigma_{j+2}^\alpha + (-1)^j J_3^{\alpha} \sigma_{j+1}^\alpha(\vec{\sigma}_{j}  \times \vec{\sigma}_{j+2} )^\alpha\big],
\end{eqnarray}
where $2N$ is the number of sites, $\sigma_j^\alpha$ is the Pauli matrix at $j$-th site along the $\alpha$-direction,
and $(\vec{\sigma}_{j}  \times \vec{\sigma}_{j+2} )^\alpha$ is the operator vector along the $\alpha$-direction.
Here NN coupling has the $XXZ$-type anisotropy. Without losing generality, we put
\begin{eqnarray}
J_1^x=J_1^y=\cosh(2a), \quad J_1^z=\cosh\eta,
\end{eqnarray}
where $a$ is a model parameter and $\eta$ is the crossing or anisotropic parameter. $J_2$ describes the NNN coupling which is isotropic.
The integrability of Hamiltonian (\ref{Ham1}) requires that the coupling constant $J_2$ satisfies
\begin{eqnarray}
J_2=-\frac{\sinh^2(2a)\cosh\eta}{2\sinh^2\eta}.\label{coeffJ2}
\end{eqnarray}
$J_3^\alpha$ describes the chiral three-spin interaction which satisfies the constraint
\begin{eqnarray}
J_3^x=J_3^y= \frac{i \sinh(2a)}{2 \sinh\eta}\cosh\eta, \quad J_3^z=\frac{i \sinh(2a)}{4 \sinh\eta}\cosh(2a).\label{coeffJ3}
\end{eqnarray}
Thus the spin chirality terms are anisotropic. The Hermitian of Hamiltonian (\ref{Ham1}) requires that $a$ is real if $\eta$ is imaginary, and $a$ is imaginary if $\eta$ is real.
If $a=0$, the NNN and three-spin chirality terms vanish and the model (\ref{Ham1}) reduces to the ordinary XXZ spin chain.
The antiperiodic boundary condition is achieved by
\begin{eqnarray}
\sigma^{\alpha}_{2N+n}=\sigma^{x}_{n} \sigma^{\alpha}_{n} \sigma^{x}_{n},\quad  n=1,2, \quad \alpha=x,y,z.\label{APB}
\end{eqnarray}

The Hamiltonian (\ref{Ham1}) is generated by the generation functionals $t(u)$ and $\hat t(u)$ as
\begin{eqnarray}\label{J1J2ham}
H = -\phi^{1-N}(2a)\sinh\eta\big\{ \hat{t}(-a)\frac{\partial \, t(u)}{\partial u}\big|_{u=a}+ \hat{t}(a) \frac{\partial \, t(u)}{\partial u}\big|_{u=-a} \big\}+E_0.
\end{eqnarray}
Here the constants $\phi(2a)$ and $E_0$ are given by
\begin{eqnarray}
\phi(2a)=-\frac{\sinh(2a+\eta)\sinh(2a-\eta)}{\sinh^2\eta},\quad E_0=-\frac{N\cosh\eta[\cosh^2(2a)-\cosh(2\eta)]}{\sinh^2\eta}.
\end{eqnarray}
$t(u)$ and $\hat t(u)$ are the transfer matrices with the definitions
\begin{eqnarray}
&&t(u)=tr_0\{\sigma^x_0 R_{0,1}(u+a) R_{0,2}(u-a) \cdots R_{0,2N-1}(u+a) R_{0,2N}(u-a)\}, \nonumber \\
&&\hat{t}(u)=tr_0 \{\sigma^x_0 R_{0,2N}(u+a) R_{0,2N-1}(u-a)\cdots R_{0,2}(u+a) R_{0,1}(u-a)\},  \label{trans}
\end{eqnarray}
where $tr_0$ means the partial trace in the auxiliary space ${\bf V}_0$, the subscript $j=\{1, \cdots, 2N\}$ denotes the $j$-th quantum or physical space ${\bf V}_j$,
$u$ is the spectral parameter and $R_{0,j}(u)$ is the six vertex $R$-matrix defined in the tensor space ${\rm\bf V}_0\otimes {\rm\bf V}_j$
\begin{eqnarray}
R_{0,j}(u)=\frac{\sinh(u+\eta)+\sinh u}{2\sinh \eta}+\frac{1}{2} (\sigma^x_0 \sigma^x_j +\sigma^y_0 \sigma^y_j) + \frac{\sinh(u+\eta)-\sinh u}{2\sinh \eta} \sigma^z_0 \sigma^z_j.
\label{R-matrix}
\end{eqnarray}
From Eq.(\ref{trans}), we know that the transfer matrices $t(u)$ and $\hat t(u)$ are defined in the tensor space ${\rm\bf V}_1\otimes {\rm\bf
V}_2\otimes\cdots\otimes{\rm\bf V}_{2N}$.

Throughout this paper, we adopt the standard notations. The ${\rm\bf V}$ denotes a $2$-dimensional linear space.
For any matrix $A\in {\rm End}({\rm\bf V})$, $A_j$ is an
embedding operator in the tensor space ${\rm\bf V}\otimes
{\rm\bf V}\otimes\cdots$, which acts as $A$ on the $j$-th space and as
identity on the other factor spaces. For any matrix $B\in {\rm End}({\rm\bf V}\otimes {\rm\bf V})$, $B_{i,j}$ is an embedding
operator of $B$ in the tensor space, which acts as identity
on the factor spaces except for the $i$-th and $j$-th ones.

The $R$-matrix (\ref{R-matrix}) has following  properties
\begin{eqnarray}
&&\hspace{-1.5cm}\mbox{ Initial
condition}:\,R_{0,j}(0)= P_{0,j},\label{Int-R} \\
&&\hspace{-1.5cm}\mbox{ Unitarity
relation}:\,R_{0,j}(u)R_{j,\,0}(-u)= \phi(u)\times{\bf id}, \label{Unitarity} \\
&&\hspace{-1.5cm}\mbox{ Crossing
relation}:\,R_{0,j}(u)=V_0R_{0,j}^{t_j}(-u-\eta)V_0,\quad V_0=-i\sigma_0^y,
\label{crosing-unitarity} \\
&&\hspace{-1.5cm}\mbox{ PT-symmetry}:\,R_{0,j}(u)=R_{j,\,0}(u)=R^{t_0\,t_j}_{0,j}(u),\label{PT} \\
&&\hspace{-1.4cm}\mbox{$Z_2$-symmetry}: \;\;
\sigma^\alpha_0\sigma^\alpha_jR_{0,j}(u)=R_{0,j}(u)\sigma^\alpha_0\sigma^\alpha_j,\quad
\mbox{for}\,\,\,
\alpha=x,y,z,\label{Z2-sym} \\
&&\hspace{-1.5cm}\mbox{ Quasi-periodicity}:\, R_{0,j}(u+i\pi)=-\sigma^z_0R_{0,j}(u)\sigma^z_0,\label{quasi-}\\
&&\hspace{-1.5cm}\mbox{ Fusion relation}:\, R_{0,j}(-\eta)=-2P^{(-)}_{0,j},\label{fu-}
\end{eqnarray}
where ${\bf id}$ means the identity operator, $R_{j,0}(u)=P_{0,j}R_{0,j}(u)P_{0,j}$ with $P_{0,j}$ being
the permutation operator, $t_l$ denotes transposition in
the $l$-th space with $l=\{0, j\}$, $P^{(-)}_{0,j}$ is the one-dimensional antisymmetric projection operator, and $P^{(-)}_{0,j}=(1- P_{0,j})/2$. Besides, the $R$-matrix (\ref{R-matrix}) satisfies the Yang-Baxter equation
\begin{eqnarray}
R_{0,j}(u_1-u_2)R_{0,l}(u_1-u_3)R_{j,l}(u_2-u_3)
=R_{j,l}(u_2-u_3)R_{0,l}(u_1-u_3)R_{0,j}(u_1-u_2).\label{QYB}
\end{eqnarray}

Using the crossing symmetry (\ref{crosing-unitarity}), we obtain the relations between transfer matrices $t(u)$ and $\hat t(u)$
\begin{eqnarray}\label{tt}
t(u)=-\hat{t}(-u-\eta), \quad \hat{t}(u)=-t(-u-\eta).
\end{eqnarray}
Meanwhile, $t(u)$ and $\hat t(u)$ have the periodicity
\begin{eqnarray}
t(u+i\pi)=(-1)^{2N-1}t(u), \quad \hat t(u+i\pi)=(-1)^{2N-1}\hat t(u).
\end{eqnarray}
From the commutation relation (\ref{Z2-sym}) and Yang-Baxter equation (\ref{QYB}),
one can prove that the transfer matrices with different spectral parameters commute with each other, i.e.,
\begin{eqnarray}
[t(u), t(v)]=[\hat t(u), \hat t(v)] =[t(u), \hat t(u)]=0. \label{t-commu1}
\end{eqnarray}
Therefore, $t(u)$ and $\hat t(u)$ serve as the generating functionals of conserved quantities of the
system. The Hamiltonian is generated by the transfer matrices as (\ref{J1J2ham}), then
the model (\ref{Ham1}) is integrable. We note that the transfer matrices $t(u)$ and $\hat t(u)$ have common eigenstates.

\section{The exact solution and $t-W$ scheme}
\setcounter{equation}{0}

Now, we exactly solve the Hamiltonian (\ref{Ham1}). The energy spectrum of the system is determined by the eigenvalues of
transfer matrices $t(u)$ and $\hat t(u)$.
From the one-to-one correspondence (\ref{tt}), we know that $t(u)$ and $\hat t(u)$ are not independent,
thus we only need to diagonalize the transfer matrix $t(u)$.
The process of diagonalizing $t(u)$ is as follows.
According to the definition (\ref{trans}), $t(u)$ is an operator-valued trigonometric polynomial of $u$ with the degree $2N-1$.
Thus the eigenvalue of $t(u)$ is a trigonometric polynomial of $u$
with the degree $2N-1$, which can be completely determined by $2N$ constraints.
Therefore, the next task is to seek these constraints. The basic technique is fusion.

Fusion is a significant method and has various applications in the representation theory of quantum algebras \cite{Kulish81,Resh83}.
The main idea of fusion is the $R$-matrix degenerates into the projection
operators at some special points. Based on it,
we can obtain the high-dimensional representation of ceratin algebraic symmetry. We shall note that
some new conserved quantities including the novel Hamiltonian quantifying some interesting interactions
could also be constructed using the fusion.

The $R$-matrix (\ref{R-matrix}) is a $4\times 4$ matrix. At the point of $u=-\eta$,
$R_{12}(u)$ degenerates into the one-dimensional antisymmetric projection operator $P_{1,2}^{(-)}$.
The accompanied three-dimensional symmetric projection operator is
$P^{(+)}_{1,2}=(1+ P_{1,2})/2$. Using the fusion, we obtain
\begin{eqnarray}
t(u)t(u-\eta)=tr_{1,2}\{P^{(-)}_{1,2}T_2(u)T_1(u-\eta)P^{(-)}_{1,2}\}+tr_{1,2}\{P^{(+)}_{1,2}T_2(u)T_1(u-\eta)P^{(+)}_{1,2}\}.\label{ai}
\end{eqnarray}
During the derivation, we have used the relations
\begin{eqnarray}
{P}_{1,2}^{(-)}+{P}_{1,2}^{(+)}={\bf id},\quad {P}_{1,2}^{(-)}{P}_{1,2}^{(+)}={P}_{1,2}^{(+)}{P}_{1,2}^{(-)}=0.
\end{eqnarray}
With the help of properties of projection operators, we obtain following $t-W$ relation
\begin{eqnarray}
t(u)t(u-\eta)=-d(u+\eta)d(u-\eta)\times{\bf id}+d(u){\bf W}(u),\label{tw}
\end{eqnarray}
where
\begin{eqnarray}
d(u)=\frac{\sinh^N(u+a)\sinh^N(u-a)}{\sinh^{2N}\eta},\label{ai1}
\end{eqnarray}
and ${\bf W}(u)$ is the undetermined operator.
It is obvious that $d(\pm a)=0$, thus the $t-W$ relation (\ref{tw}) is closed at the points of $u=\pm a$, i.e.,
\begin{eqnarray}
t(a)t(a-\eta)=-d(a+\eta)d(a-\eta)=t(-a)t(-a-\eta).
\end{eqnarray}

The fusion does not break the integrability. Thus $t(u)$ and ${\bf W}(u)$ commute with each other and they have the common eigenstates.
Assume $|\Psi\rangle$ is a common eigenstate
\begin{eqnarray}
t(u)|\Psi\rangle=\Lambda(u)|\Psi\rangle, \quad {\bf W}(u)|\Psi\rangle=W(u)|\Psi\rangle,
\end{eqnarray}
where $\Lambda(u)$ and $W(u)$ are the eigenvalues of $t(u)$ and ${\bf W}(u)$, respectively.
Acting Eq.(\ref{tw}) on the eigenstate $|\Psi\rangle$, we have
\begin{eqnarray}
\Lambda(u)\Lambda(u-\eta)=-d(u+\eta)d(u-\eta)+d(u)W(u).\label{t-w}
\end{eqnarray}
At the points of $u=\pm a$, the relation (\ref{t-w}) reduces to
\begin{eqnarray}
\Lambda(a)\Lambda(a-\eta)=-d(a+\eta)d(a-\eta)=\Lambda(-a)\Lambda(-a-\eta).
\end{eqnarray}

Because the eigenstate $|\Psi\rangle$ does not depend upon $u$, the eigenvalue $\Lambda(u)$ is a trigonometric polynomial of $u$ with the degree $2N-1$.
Meanwhile, the eigenvalue should satisfy the periodicity $\Lambda(u+i\pi)=(-1)^{2N-1}\Lambda(u)$.
Thus, we parameterize the eigenvalue $\Lambda(u)$ as
\begin{eqnarray}
\Lambda(u)=\Lambda_0\,\prod_{j=1}^{2N-1}\,\sinh (u-z_j+\eta/2),\label{Zero-points}
\end{eqnarray}
where $\{z_j-\eta/2|j=1,\cdots,2N-1\}$ are the $2N-1$ zero roots
and $\Lambda_0$  is an overall coefficient.
From Eqs.(\ref{ai1}) and (\ref{t-w}), we also know that the eigenvalue ${W}(u)$ is a trigonometric polynomial of $u$ with the degree $2N$.
We parameterize ${W}(u)$ as
\begin{eqnarray}
W(u)=W_0\sinh^{-2N}\eta\,\prod_{l=1}^{2N}\sinh(u-w_l),
\end{eqnarray}
where $\{w_j |j=1,\cdots,2N\}$ are the $2N$ zero roots and
$W_0$ is a constant.

The polynomial analysis shows that (\ref{t-w}) is a polynomial equation of $e^u$ with the degree $4N$, thus gives $4N+1$ independent constraints for the coefficients,
which are sufficient to completely determine the $2N-1$ shifted zero roots $\{z_j\}$, $2N$ zero roots $\{w_l\}$, the constants $\Lambda_0$ and $W_0$.
Since $\Lambda(u)$ is a trigonometric polynomial of $u$ with the degree $2N-1$,
the leading terms in the right hand side of Eq.(\ref{t-w}) must be zero. Then we have
\begin{eqnarray}
W_0e^{\pm\sum_{l=1}^{2N} w_l}=1.\label{BA0}
\end{eqnarray}
Putting $u=\{z_j-\eta/2| j=1,\cdots, 2N-1\}$ in Eq.(\ref{t-w}), we obtain the first set of constraints of zero roots $\{z_j\}$ and $\{w_l\}$
\begin{eqnarray}
&&\sinh^N(z_j+\frac\eta2+a)\,\sinh^N(z_j+\frac\eta2-a)\sinh^N(z_j-\frac{3\eta}2+a)\,\sinh^N(z_j-\frac{3\eta}2-a)\nonumber\\
&&=W_0\,\sinh^N (z_j-\frac\eta2+a)\,\sinh^N (z_j-\frac\eta2-a)\prod_{l=1}^{2N}\sinh(z_j-\frac\eta2-w_l),\nonumber\\
&& \quad j=1,\cdots, 2N-1.\label{BA1}
\end{eqnarray}
Putting $u=\{w_l|l=1, \cdots, 2N\}$ in Eq.(\ref{t-w}), we obtain the second set of constraints of zero roots $\{z_j\}$ and $\{w_l\}$
\begin{eqnarray}
&&-\sinh^{-4N}\eta\,\sinh^N(w_l+\eta+a)\,\sinh^N(w_l+\eta-a)\sinh^N(w_l-\eta+a)\,\sinh^N(w_l-\eta-a)\nonumber\\
&&=\Lambda_0^2\prod_{j=1}^{2N-1}\sinh(w_l-z_j+\frac\eta2)\sinh(w_l-z_j-\frac\eta2),\quad l=1,\cdots, 2N.\label{BA2}
\end{eqnarray}
The coefficient $\Lambda_0$ can be determined by putting $u=a$ in Eq.(\ref{t-w}) as
\begin{eqnarray}
&&\Lambda_0^2\prod_{j=1}^{2N-1}\sinh(a-z_j+\frac\eta2)\sinh(a-z_j-\frac\eta2)\nonumber\\
&&=(-1)^{N-1}\sinh^{-2N}\eta\,\sinh^{N}(2a+\eta)\,\sinh^N(2a-\eta).\label{BA3}
\end{eqnarray}
Then the $4N+1$ parameters $\Lambda_0$, $W_0$, $\{z_j\}$ and $\{w_l\}$ should satisfy the $4N+1$ BAEs (\ref{BA0})-(\ref{BA3}).

According to the construction (\ref{J1J2ham}), the energy spectrum of the Hamiltonian (\ref{Ham1}) can be expressed in terms of the zero roots $\{z_j\}$ as
\begin{eqnarray}\label{J1J2redu}
&&E=\phi^{1-N}(2a)\sinh\eta \big\{ \Lambda(a-\eta)\frac{\partial \Lambda(u)}{\partial u}\big|_{u=a}
 + \Lambda(-a-\eta) \frac{\partial \Lambda(u)}{\partial u}\big|_{u=-a}\big\}+E_0\nonumber \\
&&\quad =-\phi(2a)\sinh\eta\sum_{j=1}^{2N-1}\big\{\coth(a-z_j+\eta/2)- \coth(a+z_j-\eta/2)\big\}+E_0.
\end{eqnarray}

\begin{table}[!h]
\centering
\caption{The solutions of BAEs (\ref{BA0})-(\ref{BA3}) and the energy spectrum of the Hamiltonian (\ref{Ham1}) with $2N=4$, $a=0.2$ and $\eta=0.6i$.
    Here $E_n$ is the eigenenergy and $n$ is the energy level. Each level is doubly degenerate and $W_0=1$.
    The energy $E_n$ calculated from the BAEs is exactly the same as that obtained from the numerical exact diagonalization of Hamiltonian (\ref{Ham1}).}\label{c1}
{ \footnotesize
\begin{tabular}{lllll}
\hline
 $ w_1 $ & $ w_2 $ & $ w_3 $ & $ w_4 $ & $ z_1 $ \\ \hline
 $-1.2826$ & $-0.2473$ & $0.2473$ & $1.2826$  &  $-0.3477$  \\
 $-1.0902$ & $-0.0812$ & $0.5857-0.9540i$ & $0.5857+0.9540i$  & $-0.2913$ \\
 $-0.5857-0.9540i$ & $-0.5857+0.9540i$ & $0.0812$ & $1.0902$  &  $-1.8173$\\
 $-0.7051$ & $0.0000-1.0960i$ & $0.0000+1.0960i$ & $0.7051$   & $-0.2105$ \\
 $-0.3577-0.8792i$ & $-0.3577+0.8792i$ & $0.3577-0.8792i$ & $0.3577+0.8792i$  &  $-0.8076-1.5708i$ \\
 $-0.8868$ & $0.1556-0.8924i$ & $0.1556+0.8924i$ & $0.5756$  & $-0.2332$ \\
 $-0.5756$ & $-0.1556-0.8924i$ & $-0.1556+0.8924i$ & $0.8868$  & $-0.1899-0.6221i$ \\
 $-0.1910-0.8162i$ & $-0.1910+0.8162i$ & $0.1910-0.8162i$ & $0.1910+0.8162i$  & $0.0000-1.5708i$ \\
\hline
$ z_2 $ & $ z_3 $ & $ \Lambda_{0}^2 $ & $ E_n $ & $ n $\\ \hline
$0.0000$ & $0.3477$ & $-399.7321$ & $-5.2630$ & $1$ \\
$0.0728$ & $1.8173$ & $-10.8865$ & $-2.9735$  &$2$\\
$-0.0728$ & $0.2913$ & $-10.8865$ & $-2.9735$ &  $3$\\
$0.0000-1.5708i$ & $0.2105$ & $105.6802$ & $-2.7257$ & $4$\\
$0.0000$ & $0.8076-1.5708i$ & $-5.9241$ & $0.9387$ &  $5$\\
$0.1899-0.6221i$ & $0.1899+0.6221i$ & $-127.8985$ & $2.9735$ &  $6$\\
$-0.1899+0.6221i$ & $0.2332$ & $-127.8983$ & $2.9735$ &  $7$\\
$0.0000-0.6474i$ & $0.0000+0.6474i$ & $22.4078$ & $7.0500$  &$8$\\
\hline
\end{tabular}}
\end{table}
\begin{table}[!h]
\centering
\caption{The solutions of BAEs (\ref{BA0})-(\ref{BA3}) and the energy spectrum of the Hamiltonian (\ref{Ham1}) with $2N=4$, $a=0.2i$ and $\eta=0.6$.
    Here $E_n$ is the eigenenergy and $n$ is the energy level. Each level is doubly degenerate and $W_0=1$.
    The energy $E_n$ calculated from the BAEs is exactly the same as that obtained from the numerical exact diagonalization of Hamiltonian (\ref{Ham1}).}\label{c2}
{\footnotesize
\begin{tabular}{lllll}
\hline
  $ w_1 $ & $ w_2 $ & $ w_3 $ & $ w_4 $ & $ z_1 $ \\ \hline
  $-1.6145-1.5708i$ & $0.0000-0.2869i$ & $0.0000+0.2869i$ & $1.6145+1.5708i$  & $0.0000-0.3690i$ \\
  $-1.3531+0.6671i$ & $0.0000-1.2799i$ & $0.0000-0.0542i$ & $1.3531+0.6671i$ & $0.0000-0.2868i$  \\
  $-1.3531-0.6671i$ & $0.0000+0.0542i$ & $0.0000+1.2799i$ & $1.3531-0.6671i$  & $0.0000-0.7901i$  \\
  $-1.4230$ & $0.0000-0.5427i$ & $0.0000+0.5427i$ & $1.4230$  & $0.0000-1.5708i$ \\
  $-0.8913-0.3487i$ & $-0.8913+0.3487i$ & $0.8913-0.3487i$ & $0.8913+0.3487i$  & $-0.9699-1.5708i$  \\
  $-0.8929-0.1418i$ & $-0.2682-1.4290i$ & $0.2682+1.7126i$ & $0.8929-0.1418i$  & $-0.6190-0.2120i$ \\
  $-0.8929+0.1418i$ & $-0.2682-1.7126i$ & $0.2682+1.4290i$ & $0.8929+0.1418i$  & $-0.6190+0.2120i$  \\
  $-0.8693-0.2039i$ & $-0.8693+0.2039i$ & $0.8693-0.2039i$ & $0.8693+0.2039i$  & $-0.6253$ \\
\hline
$ z_2 $ & $ z_3 $ & $ \Lambda_{0}^2 $ & $ E_n $ & $ n $\\ \hline
$0.0000$ & $0.0000+0.3690i$ & $308.1505$ & $-4.6408$  & $1$\\
$0.0000+0.0800i$ & $0.0000+0.7901i$ & $140.6619$ & $-3.3343$ & $2$ \\
$0.0000-0.0800i$ & $0.0000+0.2868i$ & $140.6615$ & $-3.3343$ & $3$ \\
$0.0000-0.1996i$ & $0.0000+0.1996i$ & $79.2054$ & $-3.1881$ & $4$ \\
$0.0000$ & $0.9699-1.5708i$ & $2.6434$ & $0.9566$ & $5$ \\
$0.0000+0.2360i$ & $0.6190-0.2120i$ & $59.4799$ & $3.3343$ & $6$ \\
$0.0000-0.2360i$ & $0.6190+0.2120i$ & $59.4799$ & $3.3343$ & $7$ \\
$0.0000-1.5708i$ & $0.6253$ & $10.2842$ & $6.8723$  & $8$\\
\hline
\end{tabular}}
\end{table}

Now, we check the above analytical results numerically. We first solve the BAEs (\ref{BA0})-(\ref{BA3})
and obtain the values of zero roots $\{z_j\}$. Substituting these values into Eq.(\ref{J1J2redu}),
we obtain the eigenenergies of the Hamiltonian (\ref{Ham1}). The results are listed in Tables \ref{c1} and \ref{c2}. After
that, we numerically diagonalize the Hamiltonian (\ref{Ham1}) with the same model parameters. We find
that the eigenvalues obtained by solving the BAEs are exactly the same as those obtained by
the exact numerical diagonalization. Therefore, the expression (\ref{J1J2redu}) gives the complete spectrum of the
system.

\section{Conserved momentum and charge operators}
\setcounter{equation}{0}

In this section, we discuss the conserved momentum and charge operators.
Although the antiperiodic boundary condition breaks the translational invariance and the $U(1)$ charge is not conserved in the present system,
we can still construct the conserved momentum and charge in the topological manifold of twisted spin spaces.

Define the shift operator $U$  as
\begin{eqnarray}\label{U}
U=\phi^{-N}(2a)t(a)t(-a).
\end{eqnarray}
With the help of transfer matrix $t(u)$ at the points of $u=\pm a$, we obtain
\begin{eqnarray}\label{tata}
&&t(a)t(-a)=\sigma_{2N}^xR_{1,2N}(2a)P_{2,2N}\cdots R_{2N-1,2N}(2a)R_{1,2}(-2a)P_{1,3}\cdots R_{1,2N}(-2a)\sigma_1^x \nonumber\\
&&\hspace{1.8cm}=\phi^{N}(2a)\sigma_{2N}^xP_{2,2N}P_{4,2N}\cdots P_{2N-2,2N} P_{1,3}P_{1,5}\cdots P_{1,2N-1}\sigma_1^x.
\end{eqnarray}
By using the properties of permutation operator, we find that $U^{2N}=1$. Accordingly, we construct the topological momentum operator as $\hat K=-i\ln U$.
Then the eigenvalues of $\hat K$ are
\begin{eqnarray}
k=\frac{\pi l}N,{~~} l=-N, -N+1,\cdots, N-1.\label{perio}
\end{eqnarray}
Substituting the parametrization (\ref{Zero-points}) of eigenvalue of $t(u)$ into Eq.(\ref{U}), we obtain that
the eigenvalue $k$ can also be expressed by the zero roots $\{z_j\}$ as
\begin{eqnarray}
k=-i\sum_{j=1}^{2N-1}\ln\frac{\sinh(a+z_j-\frac\eta2)}{\sinh(a-z_j-\frac\eta2)}{~~}mod\,\{2\pi\}. \label{ioppoi}
\end{eqnarray}

The transfer matrix $t(u)$ is the generating functional of the system. Expand $t(u)$ with respect to $u$. All the expansion coefficients
commute with each other and can be regarded as the conserved quantities. Here we consider the larding term.
From the asymptotic behavior of $t(u)$, we define the conserved charge operator $Q$ as
\begin{eqnarray}
Q=\frac{(2\sinh\eta)^{2N-1}}{4e^{{(2N-1)\eta}/2}}\lim_{u\to\infty}e^{-(2N-1)u}t(u)=\frac14(Q^++Q^-),\label{ai13}
\end{eqnarray}
where
\begin{eqnarray}
Q^\pm=\sum_{j=1}^{2N} e^{(-1)^ja}e^{\mp\frac\eta2\sum_{k=1}^{j-1}\sigma_k^z}\sigma_j^\pm e^{\pm\frac\eta2\sum_{k=j+1}^{2N}\sigma_k^z}.\label{ai3}
\end{eqnarray}
Correspondingly, the asymptotic behavior of $\Lambda(u)$ gives that the
eigenvalue of conserved charge $Q$ is
\begin{eqnarray}
q=\frac14\sinh^{2N-1}\eta\,\Lambda_0\,e^{-\sum_{k=1}^{2N-1}z_k}.
\end{eqnarray}
Some remarks are in order. In the limit of $a\to 0$, the model (\ref{Ham1}) degenerates to the antiperiodic XXZ spin chain and the factor $e^{(-1)^ja}$ in Eq.(\ref{ai3}) tends to one.
The conserved charge $Q$ (\ref{ai13}) covers the previous one given in Ref.\cite{Qiao2020}. Moreover, when $a\to 0$ and $\eta\to 0$,
the model (\ref{Ham1}) degenerates to the antiperiodic isotropic
spin chain and the $U(1)$ symmetry is recovered. In this case, the conserved charge reads $Q=\sum_{j=1}^N \sigma_j^x/2$, which is the total spin along the $x$-direction.

\section{The ground state}
\setcounter{equation}{0}

Now, we study the ground state of the system (\ref{Ham1}). For simplicity, we consider the case that $a$ is real and $\eta$ is imaginary.
By using the crossing symmetry (\ref{crosing-unitarity}),
the transfer matrix $t(u)$ can be rewritten as
\begin{eqnarray}\label{ai111}
&&t(u)=(-1)^{2N-1}tr_0\{\sigma^x_0 R_{0,1}^{t_{0}}(-u-a-\eta) R_{0,2}^{t_{0}}(-u+a-\eta) \cdots \nonumber \\
&&\qquad\quad \times R_{0,2N-1}^{t_{0}}(-u-a-\eta) R_{0,2N}^{t_{0}}(-u+a-\eta)\}.
\end{eqnarray}
Because $\eta$ is imaginary, the $R$-matrix (\ref{R-matrix}) satisfies
\begin{eqnarray}\label{R-R-relation}
	R_{0,j}^{\ast t_{j}}(-u-\eta)=R_{0,j}^{t_{0}}(u^{\ast}-\eta).
\end{eqnarray}
Substituting Eq.(\ref{R-R-relation}) into (\ref{ai111}) and taking the Hermitian conjugate, we obtain
\begin{eqnarray}\label{t-t-relation}
t^{\dagger}(u)=-t(u^{\ast}-\eta).
\end{eqnarray}
By using the $t-W$ relation (\ref{tw}), we have
\begin{eqnarray}
\Lambda(u)=- \Lambda^*(u^*-\eta), \quad W(u)=W^*(u^*),
\end{eqnarray}
which gives that both the zero roots $\{z_j\}$ and the $\{w_l\}$ take the real values or form the conjugate pairs.
These patterns allow us to calculate the physical quantities in the thermodynamic limit.

\begin{figure}[t]
\begin{center}
\includegraphics[width=5cm]{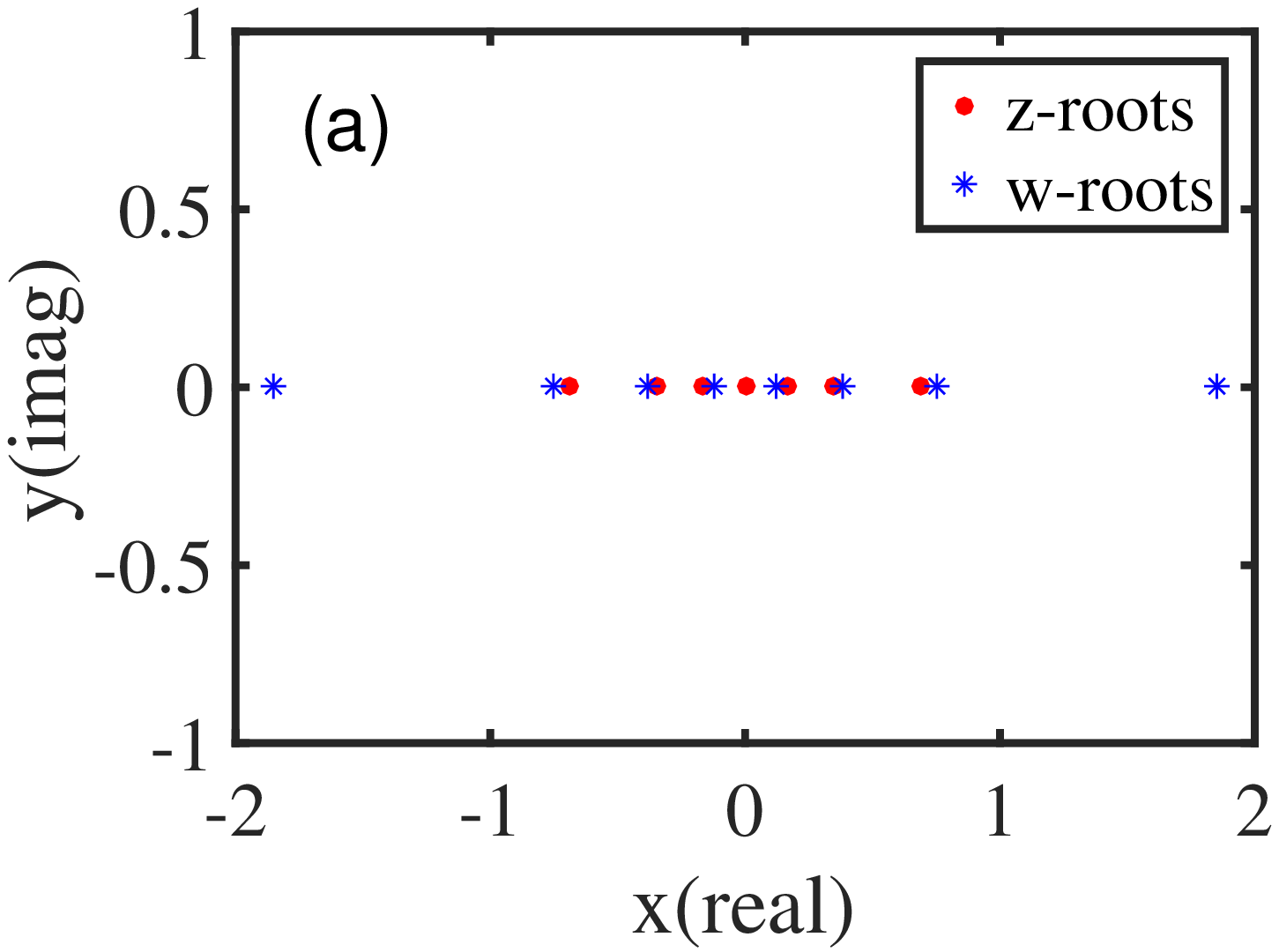}
\includegraphics[width=5cm]{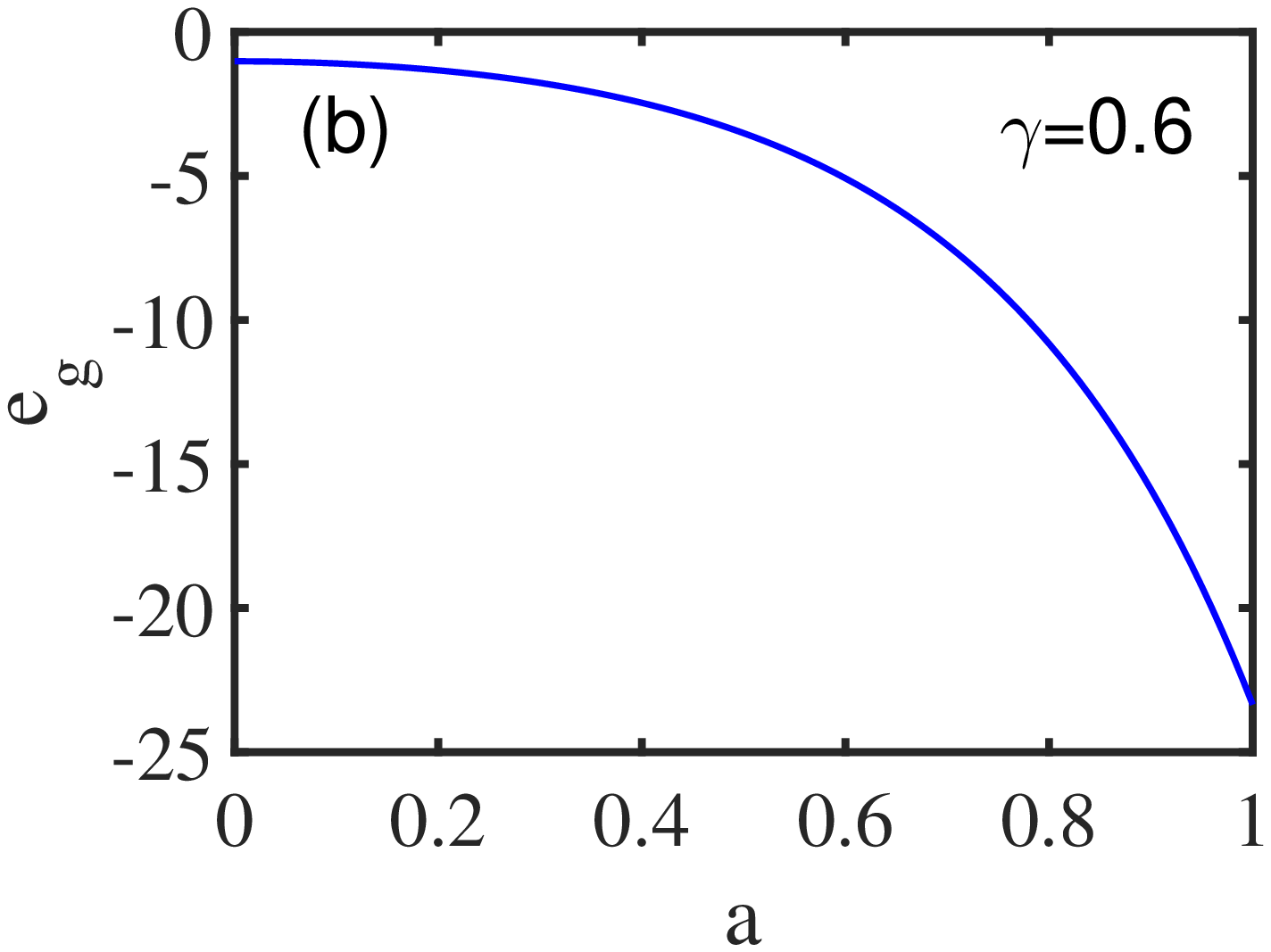}
\includegraphics[width=5cm]{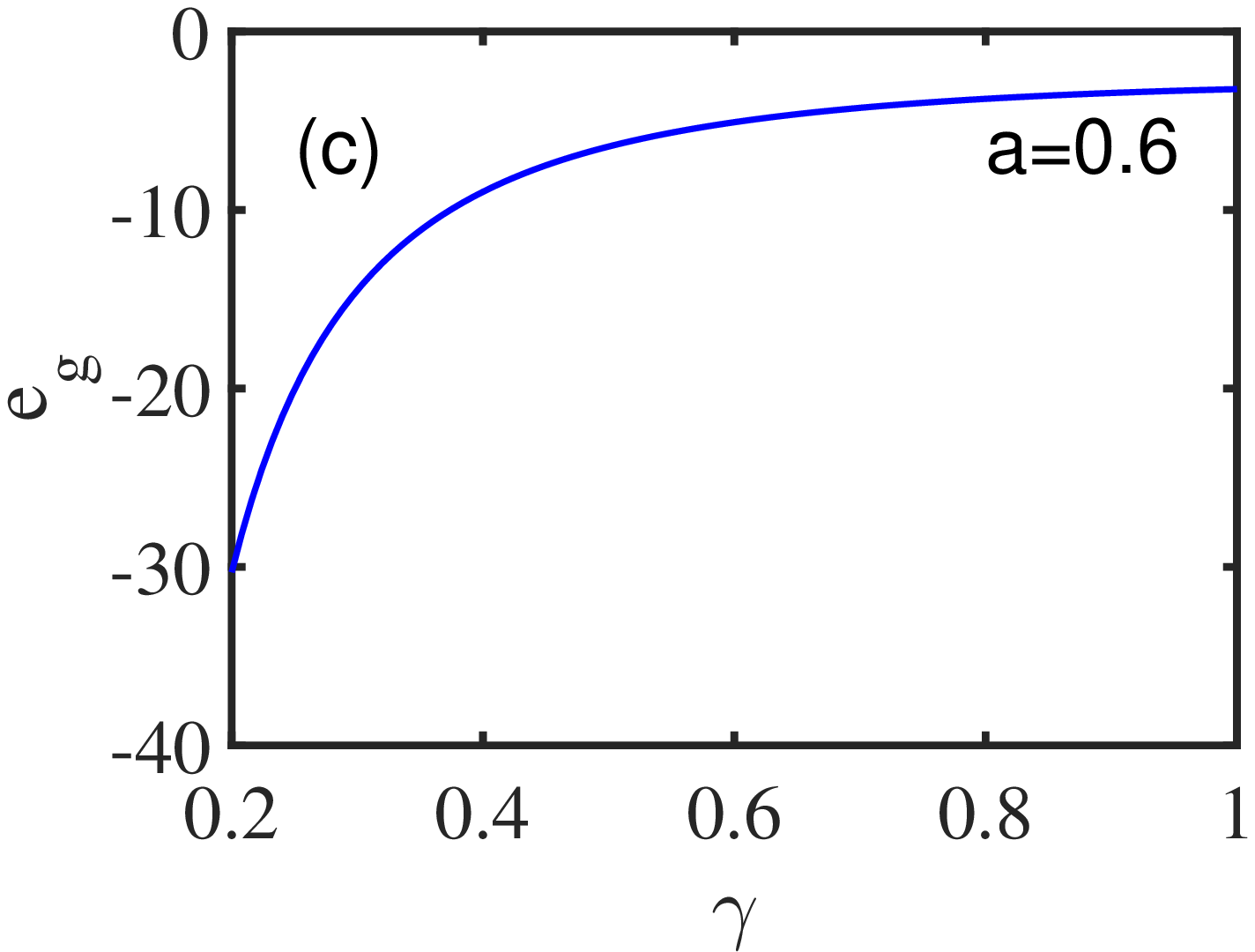}
\caption{(a) The distributions of zero roots $\{z_j\}$ and $\{w_l\}$ at the ground states obtained by solving BAEs (\ref{BA0})-(\ref{BA3}) with $2N=8$, $a=0.2$ and $\gamma=0.6$.
The ground state energy densities (\ref{ttaa}) versus the model parameter (b) $a$ with $\gamma=0.6$ and (c) $\gamma$ with $a=0.6$.}\label{fig-e1}
\end{center}
\end{figure}

At the ground state, all the zero roots $\{z_j\}$ and $\{w_l\}$ are real and distribute around the origin symmetrically.
The numerical check is shown in Fig.\ref{fig-e1}(a).
Taking the complex conjugate of BAEs (\ref{BA1}), dividing it by Eq.(\ref{BA1}) and taking the
logarithm of resulted equation, we obtain
\begin{eqnarray}
&&2\alpha_1(z_j+a)+2\alpha_1(z_j-a)-\alpha_3(z_j+a)-\alpha_3(z_j-a)=\frac{4\pi I_j}N-\frac1N\sum_{l=1}^{2N}\alpha_1(z_j-w_l),\nonumber \\
&& \qquad j=1,\cdots, 2N-1,
\label{bae1}
\end{eqnarray}
where $\alpha_n(x)=-i\ln\sinh(x-n\eta/2)+i\ln\sinh(x+n\eta/2)$ and $I_j$ is the quantum number characterizing the ground state
\begin{eqnarray*}
\{I_j\}=\left\{-N+1,-N+2,\cdots,N-2, N-1\right\}.
\end{eqnarray*}
Multiplying the complex conjugate of BAEs (\ref{BA1}) by (\ref{BA1}) and taking the
logarithm of resulted equation, we have
\begin{eqnarray}
\beta_3(z_j+a)+\beta_3(z_j-a)=\frac1N\sum_{l=1}^{2N}\beta_1(z_j-w_l),\quad j=1,\cdots, 2N-1,\label{bae2}
\end{eqnarray}
where $\beta_n(x)=\ln\sinh(x-n\eta/2)+\ln\sinh(x+n\eta/2)$.

In the thermodynamic limit $N \to\infty$, the zero roots $z_j$ and $w_l$ in Eqs.(\ref{bae1})-(\ref{bae2}) become continue variables $z$ and $w$, respectively,
and the associated functions become the continue ones. Taking the derivative of Eqs.(\ref{bae1})-(\ref{bae2}), we have
\begin{eqnarray}
&&2a_1(z+a)+2a_1(z-a)-a_3(z+a)-a_3(z-a)=4\rho(z)+4\rho^h(z)-2a_1*\sigma(z),\label{z1}\\
&&b_3(z+a)+b_3(z-a)= 2b_1*\sigma(z),\label{z2}
\end{eqnarray}
where $\gamma=-i\eta$ is real, $a_n(x)=\sin(n\gamma)/[\pi(\cosh2x-\cos n\gamma)]$, $b_n(x)=\sinh(2x)/[\pi(\cosh2x-\cos n\gamma)]$,
$\rho(z)$, $\sigma(w)$ and $\rho^h(z)$ are the densities of $z$-roots, $w$-roots and holes in the $z$-axis, respectively,
and the notation $*$ indicates the convolution.

The reason for existing the density of holes $\rho^h(z)$ at the ground state is as follows.
In the antiperiodic boundary condition, the total number of $z$-roots is $2N-1$ while there are $2N$ possible occupations in the Brillouin zone.
At the ground state, the holes should be put at the infinity of spectral space to minimize the energy.
We shall note that both $\rho(z)$ and $\rho^h(z)$ are distributed symmetrically around the origin. Thus we have $2N\int_L^\infty\rho^h(z)dz=1/2$ and $2N\int_{-\infty}^{-L}\rho^h(z)dz=1/2$, where $L\to\infty$.
This distribution feature gives that two half holes are related to two zero modes, where the energies of holes are zero in the thermodynamic limit.
Thus the ground state of the system (\ref{Ham1}) has two zero modes which carry zero energy due to the double degeneracy.

Taking the Fourier transformation of Eqs.(\ref{z1})-(\ref{z2}), we obtain
\begin{eqnarray}
\rho(z)+\rho^h(z)=\frac{\sin\frac{\pi\gamma}{2\pi-2\gamma}}{\pi-\gamma} \bigg \{ \frac{\cosh\frac{\pi z+\pi a}{\pi-\gamma}}{\cosh\frac{2\pi z+2\pi a}{\pi-\gamma} -\cos\frac{\pi\gamma}{\pi-\gamma}}+\frac{\cosh\frac{\pi z-\pi a}{\pi-\gamma}}{\cosh\frac{2\pi z-2\pi a}{\pi-\gamma} -\cos\frac{\pi\gamma}{\pi-\gamma}} \bigg \}.\label{z-density}
\end{eqnarray}
Eq.\eqref{z-density} gives the density of $z$-roots at the ground state of the system (\ref{Ham1}) in the thermodynamic limit. Based on it, the physical quantities can be calculated analytically. Substituting Eq.\eqref{z-density} into \eqref{ioppoi},
we obtain the momentum at the ground state and the result is zero. Substituting \eqref{z-density} into (\ref{J1J2redu}), we obtain the ground state energy.
Dividing it by the system size, we obtain the ground state energy density as
\begin{eqnarray}
&&e_g=\frac{\cos(2\gamma)-\cosh(4a)}{\sin\gamma}  \int
  \frac{\cos^2(2a\tau)\cosh[(\pi-2\gamma)\tau]
  \tanh[(\pi-\gamma)\tau]} {\sinh(\pi\tau)}d\tau \nonumber\\
  &&\qquad+\frac{\cos\gamma[\cosh^2(2a)-\cos(2\gamma)]}{2\sin^2\gamma}.\label{ttaa}
\end{eqnarray}
The ground state energy density (\ref{ttaa}) versus the model parameter $a$ and $\gamma$ are shown in Fig.\ref{fig-e1}(b) and (c), respectively.

\section{Elementary excitation}
\setcounter{equation}{0}

\begin{figure}[t]
\begin{center}
\includegraphics[width=7cm]{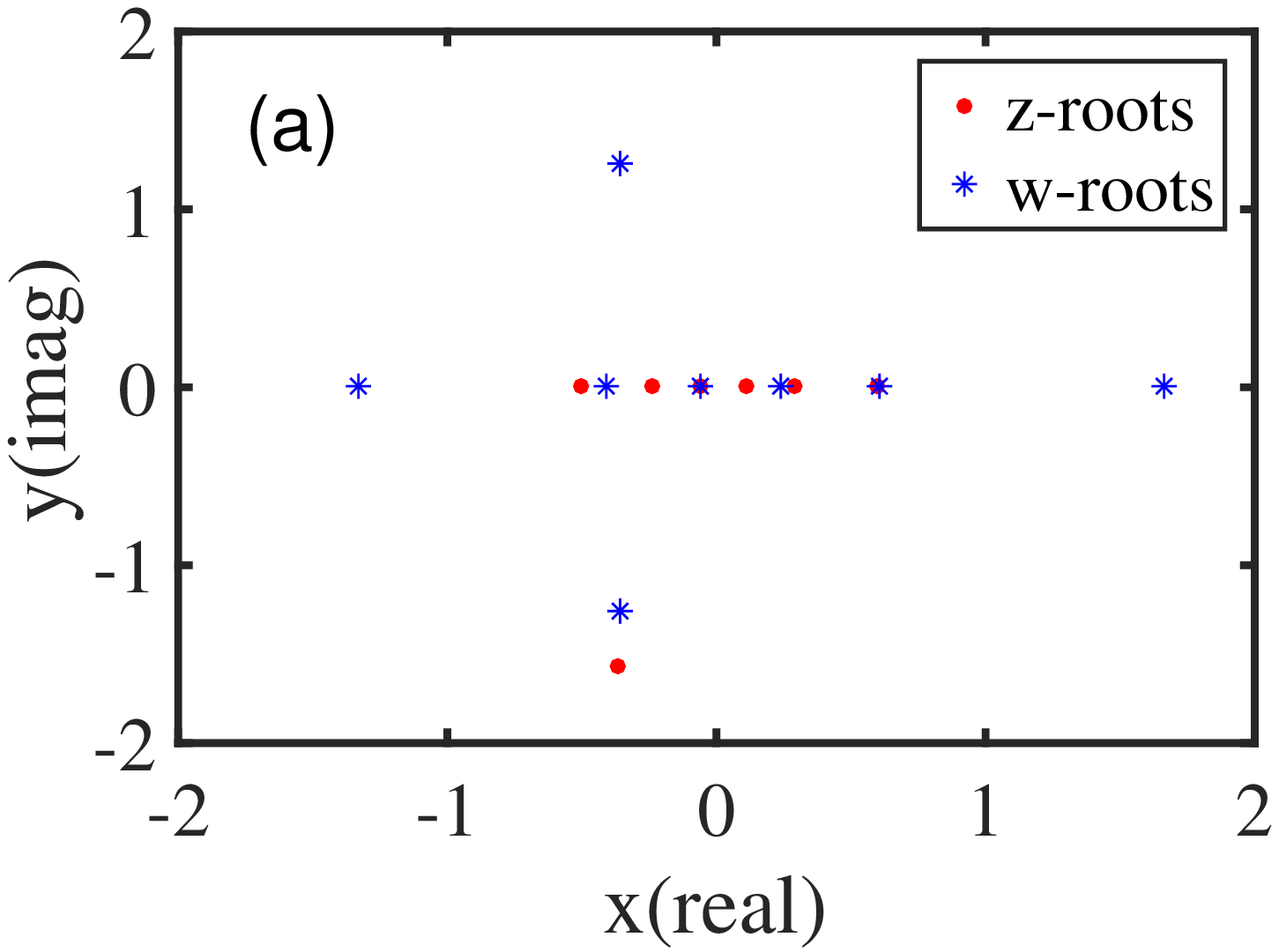}
\includegraphics[width=7cm]{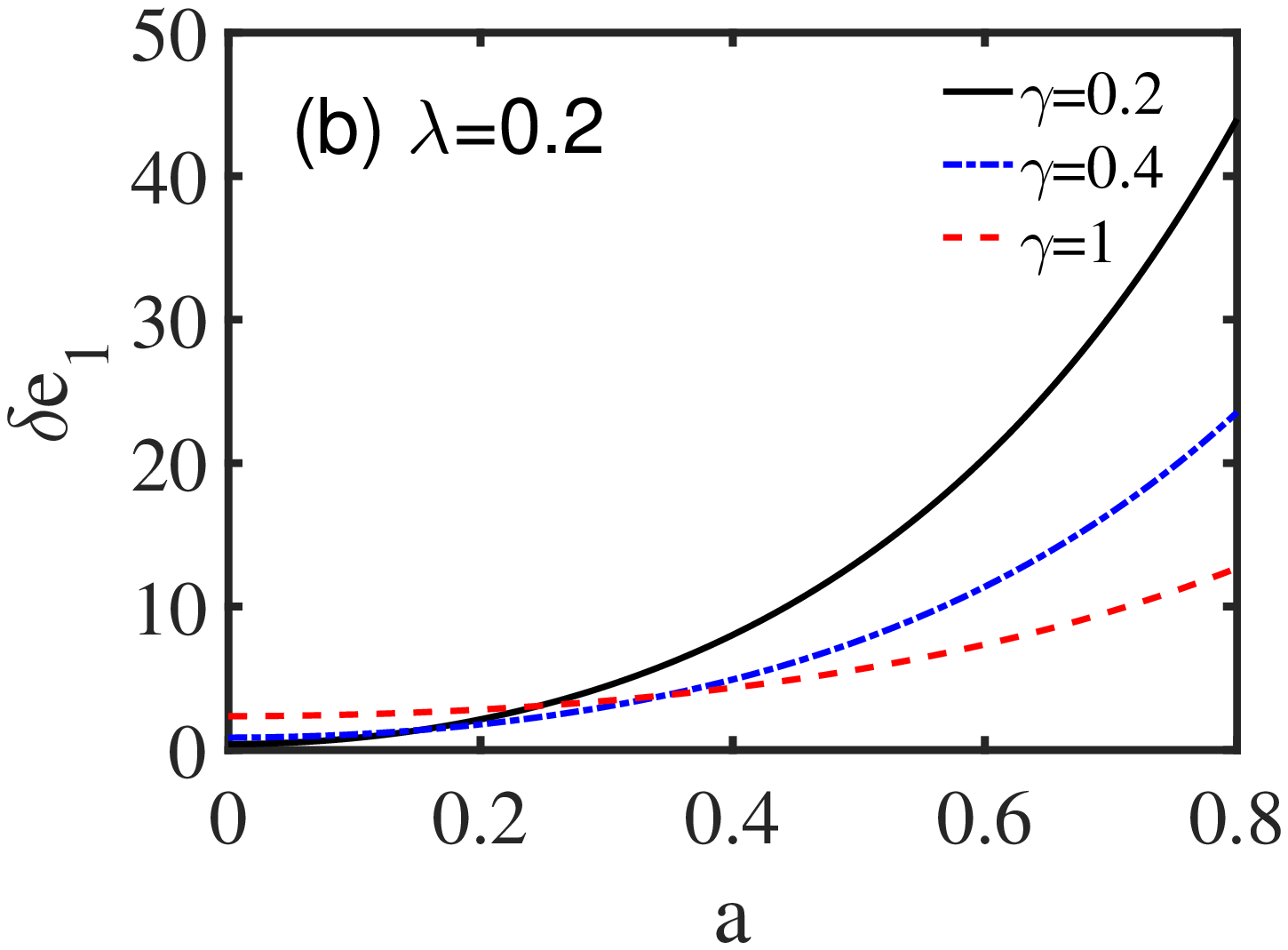}\\
\includegraphics[width=7cm]{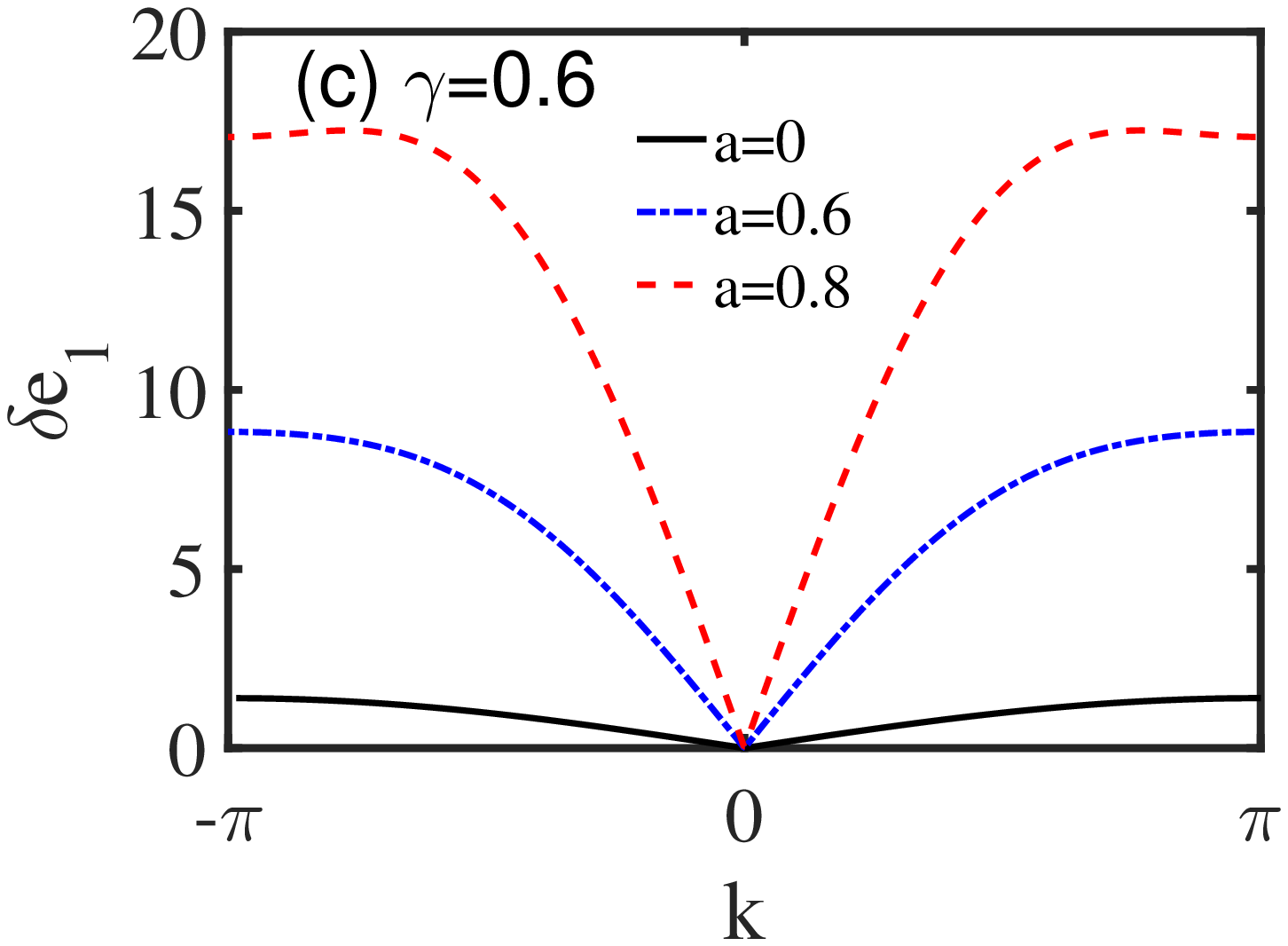}
\includegraphics[width=7cm]{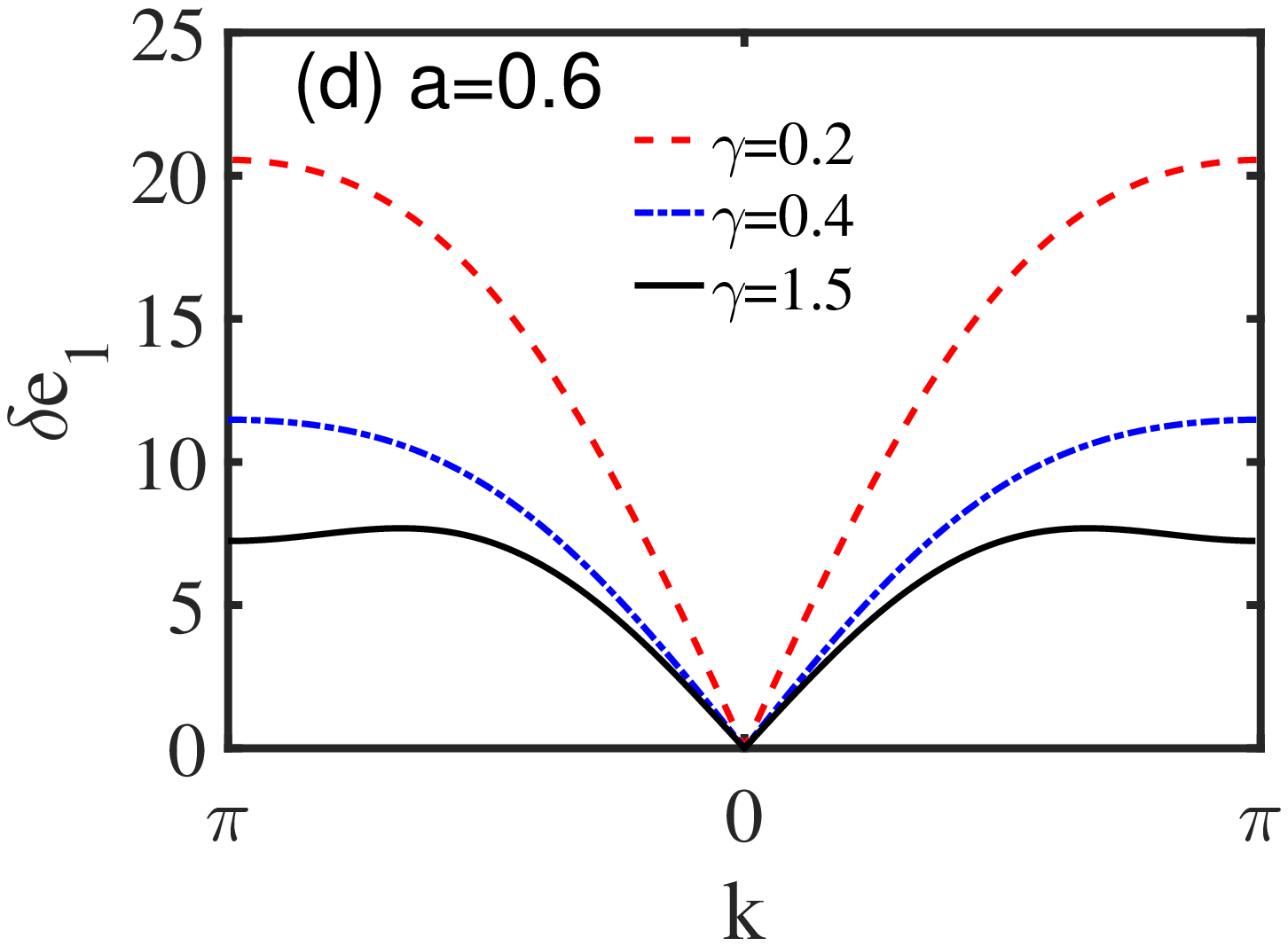}
\caption{(a) The distribution of $z$- and $w$-roots of type I elementary excitations with $2N=8$, $a=0.2$ and $\gamma=0.6$.
(b) The excited energies (\ref{e1}) versus the different values of model parameter $a$ with $\lambda=0.2$ and $\gamma=0.2, 0.4, 1$.
(c) The dispersion relation of type I elementary excitation with $\gamma=0.6$ and $a=0, 0.6, 0.8$.
(d) The dispersion relation of type I elementary excitation with $a=0.6$ and $\gamma=0.2, 0.4, 1.5$.
We shall note that the data in (b)-(d) are the results in the thermodynamic limit.}\label{fig-ee1}
\end{center}
\end{figure}

Next, we study the elementary excitation of the system.
The simplest elementary excitation is described by $(2N-2)$ real $z$-roots and one single complex root $\lambda-i \pi/2$, where $\lambda$ is a real parameter.
Accordingly, the $2N-2$ $w$-roots still remain in the real axis while two $w$-roots form a conjugate pair $\lambda_1\pm i m\gamma/2$, where $\lambda_1$ and $m$ are real parameters.
Due to the constraints of BAEs  (\ref{BA0})-(\ref{BA3}), the values of $\lambda$, $\lambda_1$ and $m$ are not independent.
The distributions of zero roots $\{z_j\}$ and $\{w_l\}$ for such an excitation with $2N=8$ are shown in Fig.\ref{fig-ee1}(a).
In the thermodynamic limit, substituting the two sets of $z$- and $w$-roots into Eqs.(\ref{z1})-(\ref{z2}), we obtain the constraints of $\lambda$, $\lambda_1$ and $m$ as
\begin{eqnarray}\label{13}
\lambda=\lambda_1,\quad m=\frac{\pi}{\gamma}-1.
\end{eqnarray}
We see that the parameter $\lambda$ is free while $m$ is fixed.
The deviation of density of $z$-roots from that at the ground state is
\begin{eqnarray}
\delta \rho_1(z)=\frac{1}{N(\gamma-\pi)}\frac{\cosh\frac{\pi (z-\lambda)}{\pi-\gamma} \cos\frac{\pi\gamma}{2\pi-2\gamma}}{\cosh\frac{2\pi (z-\lambda)}{\pi-\gamma}+\cos\frac{\pi\gamma}{\pi-\gamma}} +\frac1{2N}\delta\big(z-\lambda+\frac{i\pi}2\big). \label{poiiop}
\end{eqnarray}
Based on it, we obtain the excitation energy
\begin{eqnarray}\label{e1}
&&\delta e_1 = \frac{\cosh(4a)-\cos(2\gamma)}{\sin\gamma}\int
  \frac{\cos(2a\tau)\cos(2\lambda\tau)\cosh(\gamma\tau)
  \tanh[(\pi-\gamma)\tau]} {\sinh(\pi\tau)}d\tau \nonumber \\
  &&\qquad +
  \frac{\cosh(4a)-\cos(2\gamma)}{2\sin\gamma} \bigg[\frac{\sin\gamma}
  {\cosh(2\lambda+2a)+\cos\gamma}+\frac{\sin\gamma}{\cosh(2\lambda-2a)+\cos\gamma}\bigg],
\end{eqnarray}
which is a function of $\lambda$. The excitation energies versus the different values of model parameters $a$ and $\gamma$ with $\lambda=0.2$ are shown in Fig.\ref{fig-ee1}(b).

Substituting Eq.(\ref{poiiop}) into the thermodynamic limit expression of (\ref{ioppoi}), we obtain the corresponding momentum as
\begin{eqnarray}\label{k1}
&&k_1 = \frac{2i}{\pi-\gamma}\int\frac{\cosh\frac{\pi (z-\lambda)}{\pi-\gamma} \cos\frac{\pi\gamma}{2\pi-2\gamma}}{\cosh\frac{2\pi (z-\lambda)}{\pi-\gamma}+\cos\frac{\pi\gamma}{\pi-\gamma}}\ln\frac{\sinh(a+z-\frac{i\gamma}2)}{\sinh(a-z-\frac{i\gamma}2)}dz\nonumber\\
    &&\qquad -i\ln\frac{\sinh(a+\lambda-\frac{i\pi}2-\frac{i\gamma}2)}{\sinh(a-\lambda+\frac{i\pi}2-\frac{i\gamma}2)} {~~}mod\,\{2\pi\}.
\end{eqnarray}
Since the excitation energy $\delta e_1$ and the quasi-momentum $k_1$ rely on the single parameter $\lambda$,
the dispersion relation of this kind of elementary excitation can be derived from Eqs.(\ref{e1})-(\ref{k1}),
and the results with different model parameters $a$ and $\gamma$ are shown in Fig.\ref{fig-ee1}(c) and (d), respectively.
From Fig.\ref{fig-ee1}(c), we see that if the model parameter $a$ is big, the excitation spectrum has two peaks where the corresponding momentums are not  $\pm\pi$.
With the decreasing of $a$, the peaks converge to the point of $\pm\pi$.
When $a=0$, that is the NNN and three-spin chirality interactions vanish, the excitation spectrum
becomes to that of antiperiodic XXZ spin chain. Fig.\ref{fig-ee1}(d) shows the excitation spectrum with different values of anisotropic parameter $\gamma$.
From them, we also know that $\delta e_1 \propto k_1$ if the momentum is small, which means that it is a linear spin-wave like excitation.

\begin{figure}[t]
\begin{center}
\includegraphics[width=7cm]{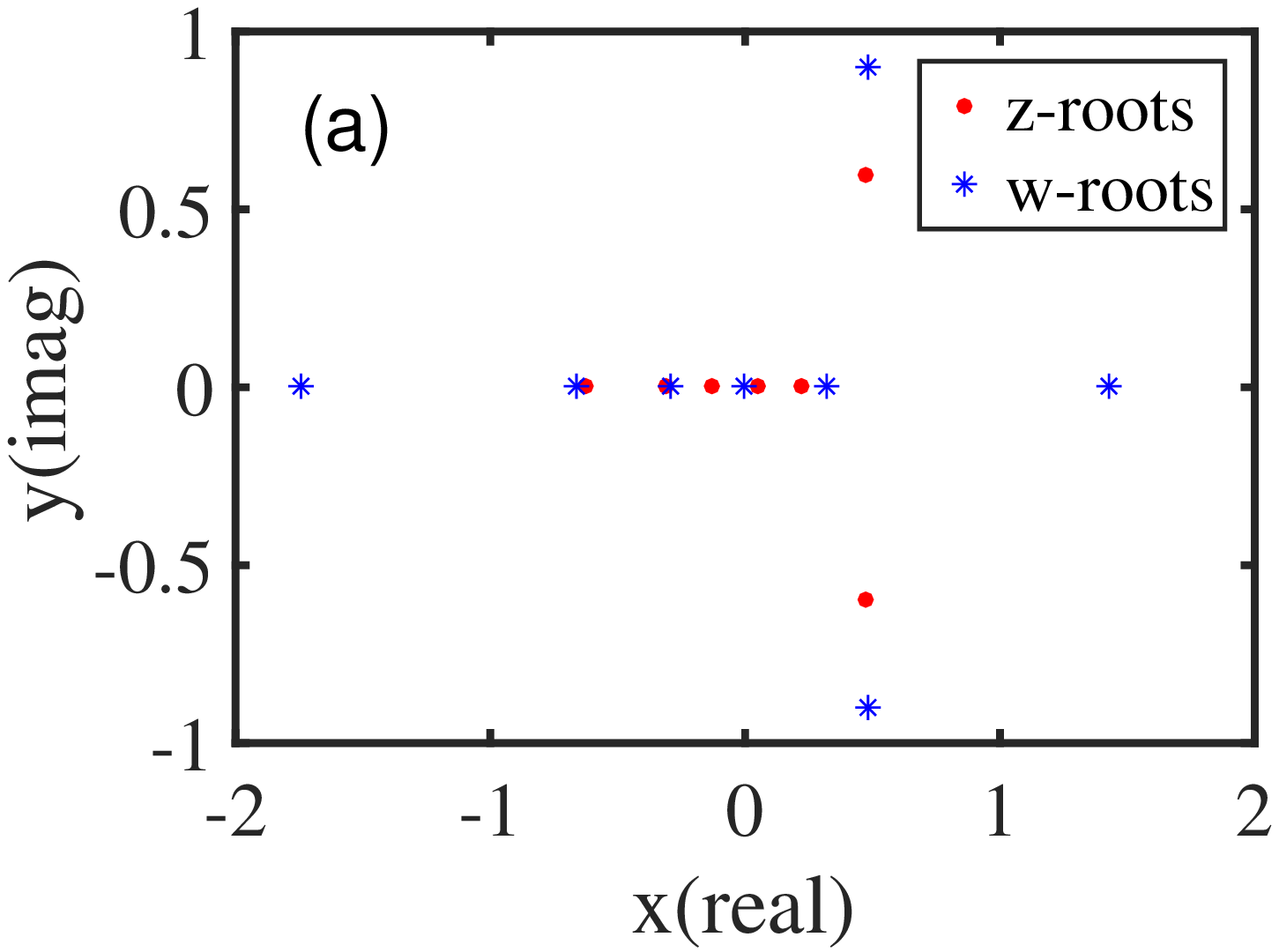}
\includegraphics[width=7cm]{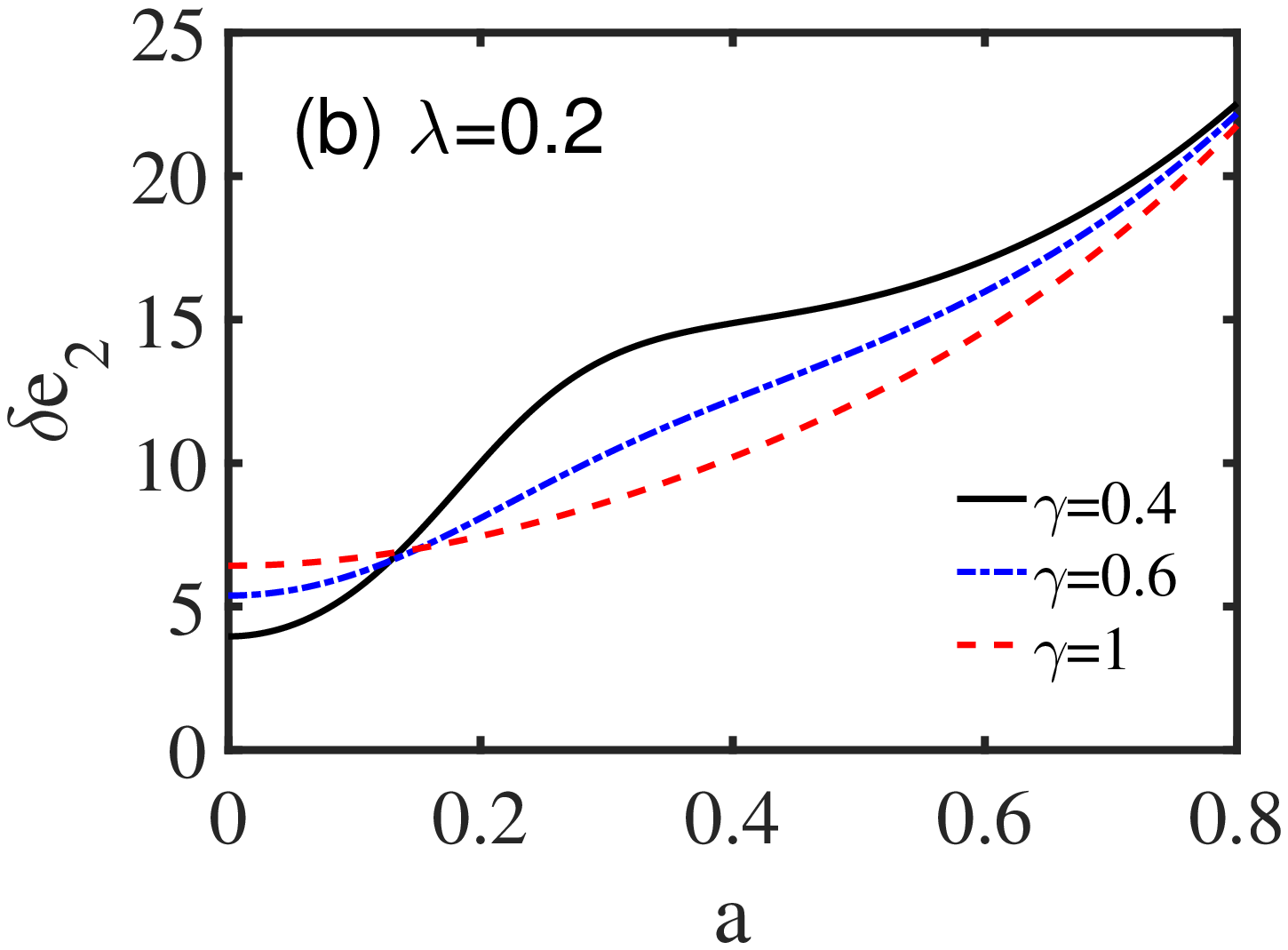}\\
\includegraphics[width=7cm]{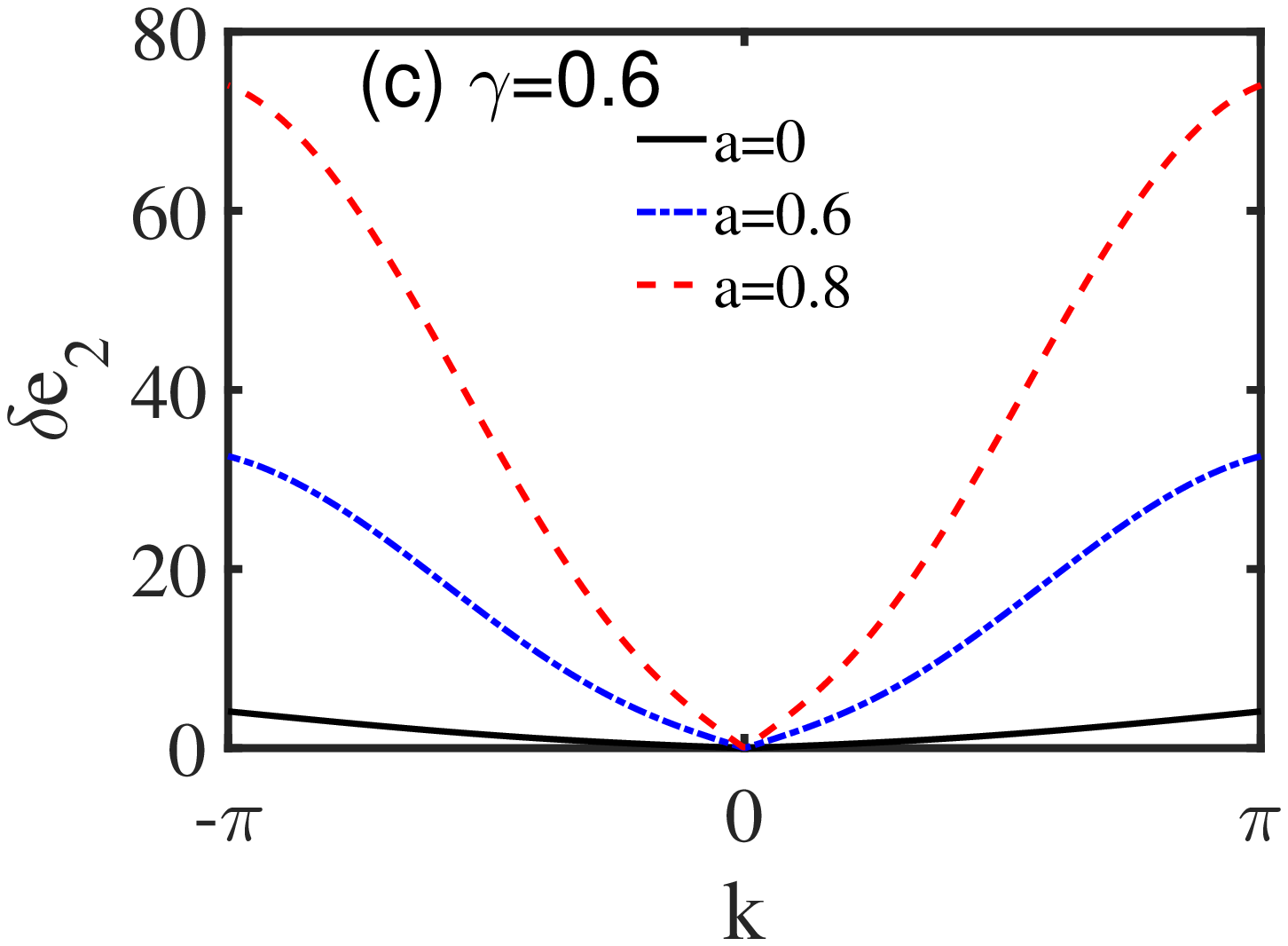}
\includegraphics[width=7cm]{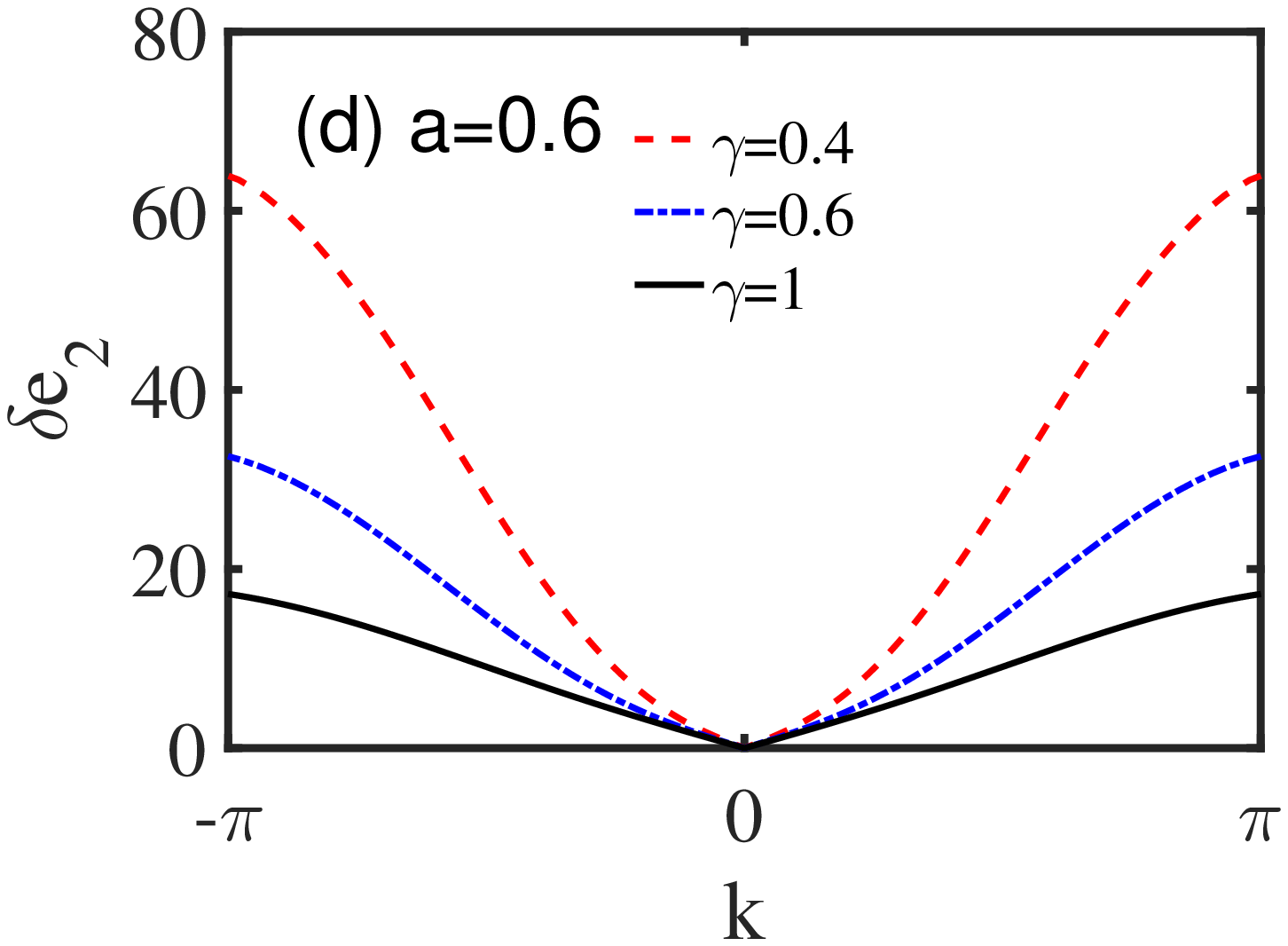}
\caption{(a) The distribution of $z$- and $w$-roots for the type II elementary excitation with $2N=8$, $a=0.2$ and $\gamma=0.6$.
(b) The excited energies (\ref{e2}) versus the different values of model parameter $a$ with $\lambda=0.2$ and $\gamma=0.4, 0.6, 1$.
(c) The dispersion relation of type II elementary excitation with $\gamma=0.6$ and $a=0, 0.6, 0.8$.
(d) The dispersion relation of type II elementary excitation with $a=0.6$ and $\gamma=0.4, 0.6, 1$. We shall note that the data in (b)-(d) are the results in the thermodynamic limit.}\label{fig-ee2}
\end{center}
\end{figure}

The second kind of elementary excitation is described by $2N-3$ real $z$-roots and one conjugate pair $\lambda\pm i \gamma$.
Using the similar procedure mentioned above, in the thermodynamic limit,
the constraints of BAEs (\ref{z1})-(\ref{z2}) give
that the $w$-roots should be $2N-2$ real values and one conjugate pair $\lambda\pm3 i \gamma/2$.
The distribution of $z$- and $w$-roots with $2N=8$ is shown in Fig.\ref{fig-ee2}(a).
The energy carried by this excitation is
\begin{eqnarray}\label{e2}
&&\delta e_2 = \frac{\cosh(4a)-\cos(2\gamma)}{\sin\gamma}
  \int
  \frac{\cos(2a\tau)\cos(2\lambda\tau)\cosh[(\pi-3\gamma)\tau]
  \tanh[(\pi-\gamma)\tau]} {\sinh(\pi\tau)}d\tau \nonumber \\ &&\qquad\;\;+
  \frac{\cosh(4a)-\cos(2\gamma)}{2\sin\gamma} \bigg[\frac{2\sin\gamma}
  {\cosh(2\lambda+2a)-\cos\gamma}+\frac{2\sin\gamma}{\cosh(2\lambda-2a)-\cos\gamma}\nonumber\\
  &&\qquad\;\;- \frac{\sin(3\gamma)}{\cosh(2\lambda+2a)-\cos(3\gamma)}
  -\frac{\sin(3\gamma)}{\cosh(2\lambda-2a)-\cos(3\gamma)}\bigg].
\end{eqnarray}
The excited energies with different values of model parameters $a$ and $\gamma$ are shown in Fig.\ref{fig-ee2}(b).
The associated quasi-momentum reads
\begin{eqnarray}\label{k2}
&&k_2 =\frac{2i}{\pi-\gamma}\int\frac{\cosh\frac{\pi (z-\lambda)}{\pi-\gamma} \cos\frac{\pi(\pi-3\gamma)}{2\pi-2\gamma}}{\cosh\frac{2\pi (z-\lambda)}{\pi-\gamma} +\cos\frac{\pi(\pi-3\gamma)}{\pi-\gamma}}\ln\frac{\sinh(a+z-\frac{i\gamma}2)}{\sinh(a-z-\frac{i\gamma}2)}dz\nonumber\\
    &&\qquad +i\ln\frac{\sinh(a+\lambda-\frac{i\gamma}2)\sinh(a-\lambda+\frac{i\gamma}2)\sinh(a-\lambda-\frac{i3\gamma}2)}
    {\sinh(a-\lambda-\frac{i\gamma}2)\sinh(a+\lambda+\frac{i\gamma}2)\sinh(a+\lambda-\frac{i3\gamma}2)} {~~}mod\,\{2\pi\}.
\end{eqnarray}
Based on Eqs.(\ref{e2}) and (\ref{k2}), the dispersion relations with different values of model parameters $a$ and $\gamma$ are shown in Fig.\ref{fig-ee2}(c) and (d), respectively.
Comparing them, we find that the contribution of parameter $a$ to the excited energy is similar to that of anisotropic parameter $\gamma$.

\begin{figure}[ht]
\begin{center}
\includegraphics[width=7cm]{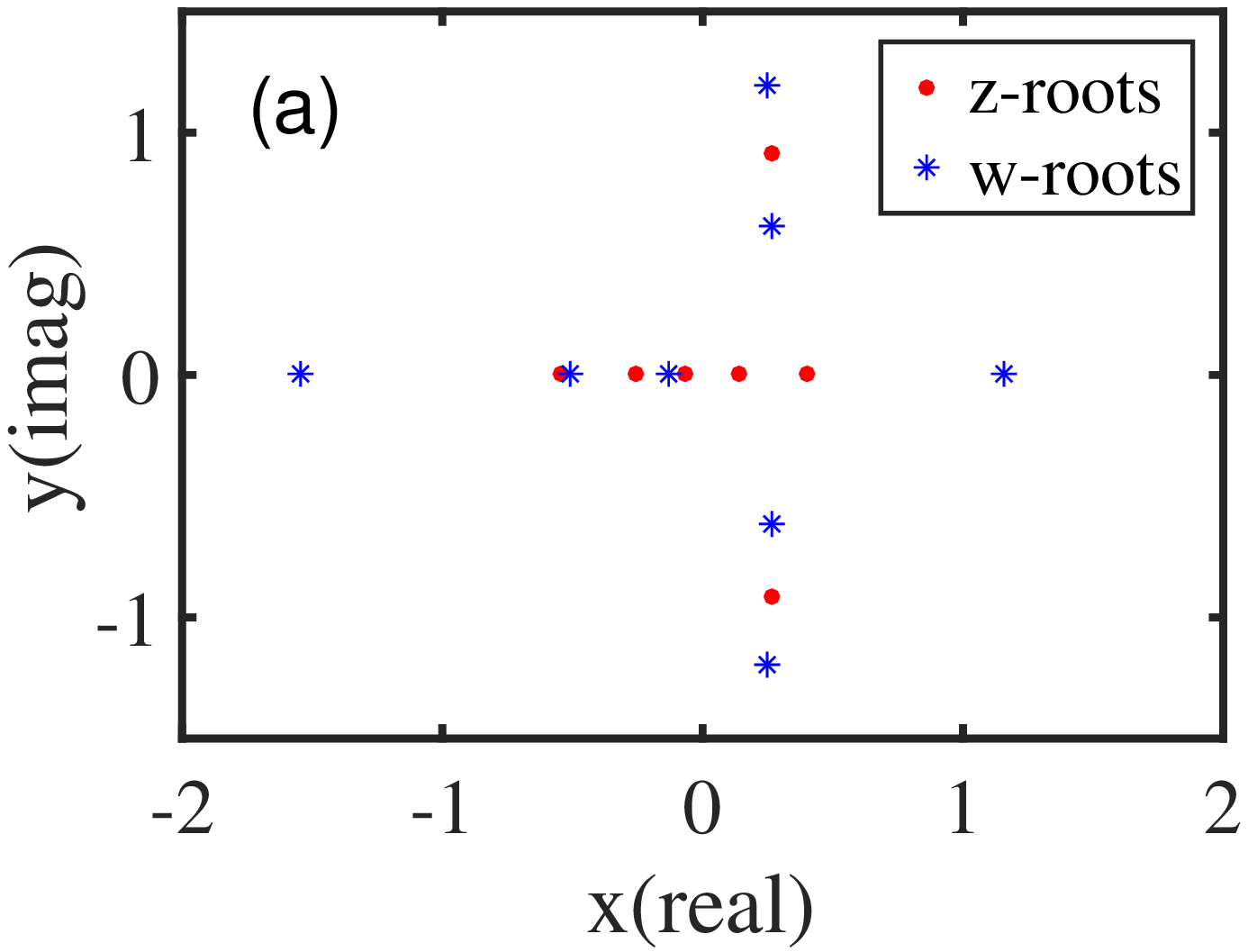}
\includegraphics[width=7cm]{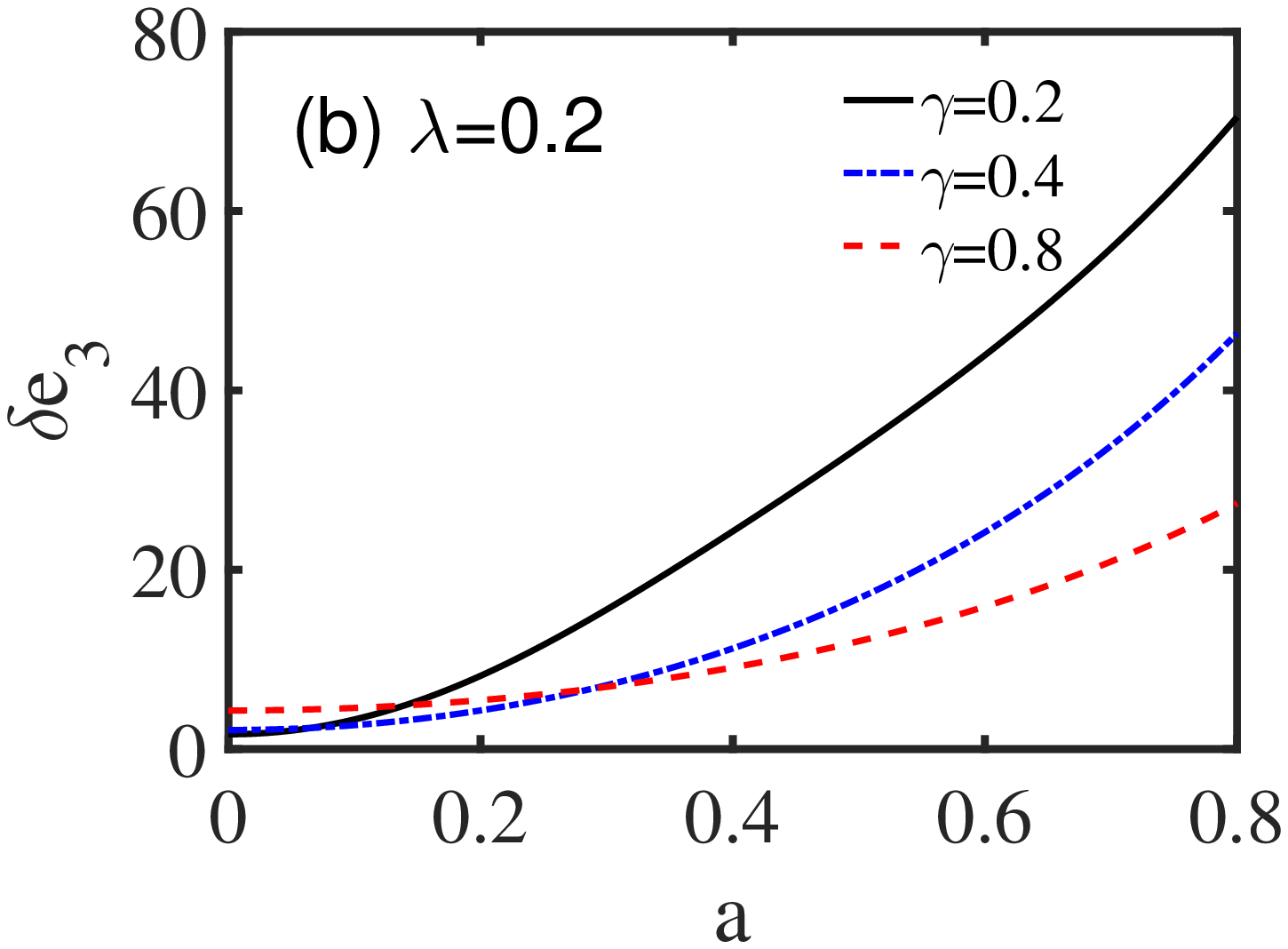}\\
\includegraphics[width=7cm]{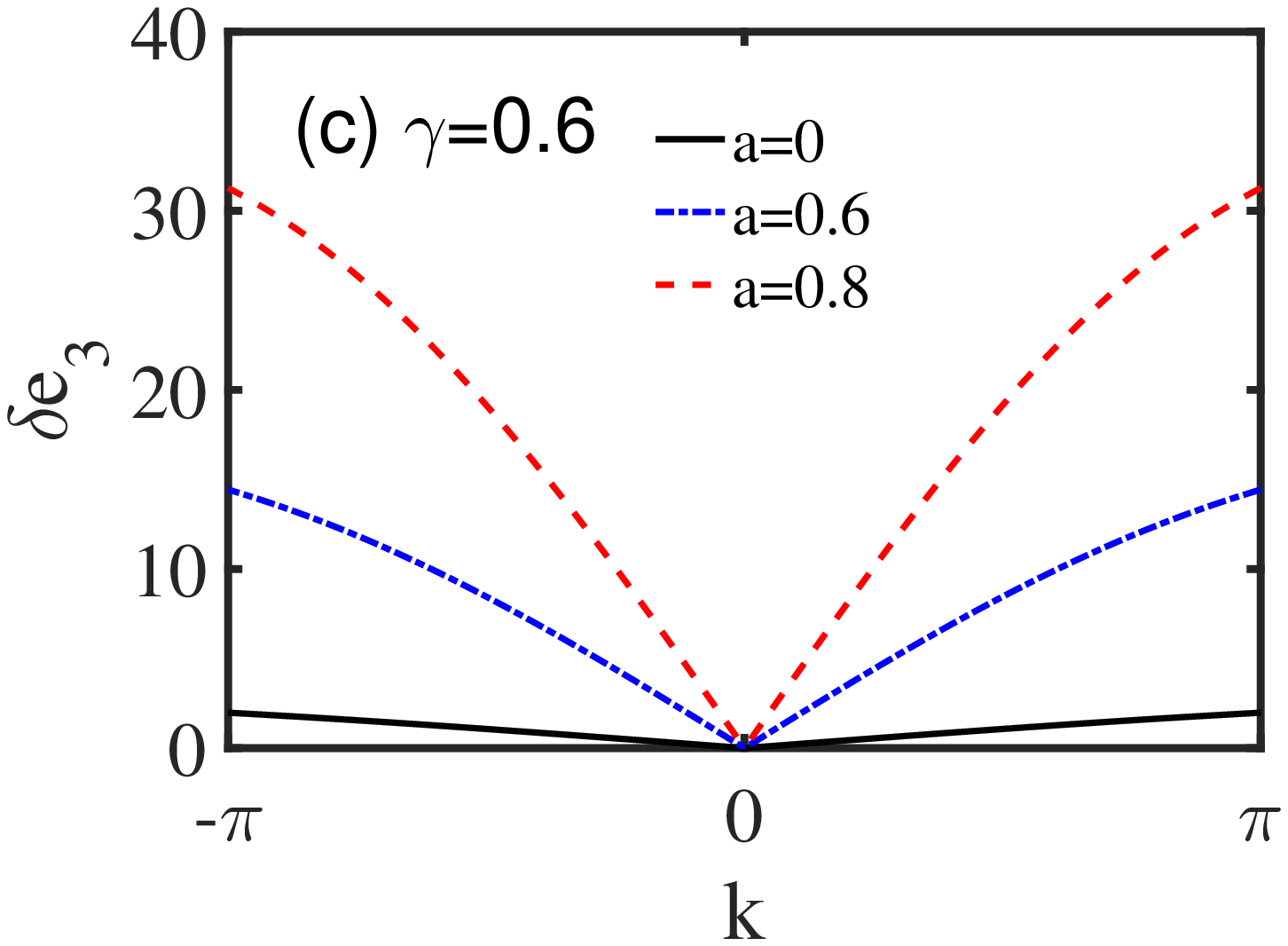}
\includegraphics[width=7cm]{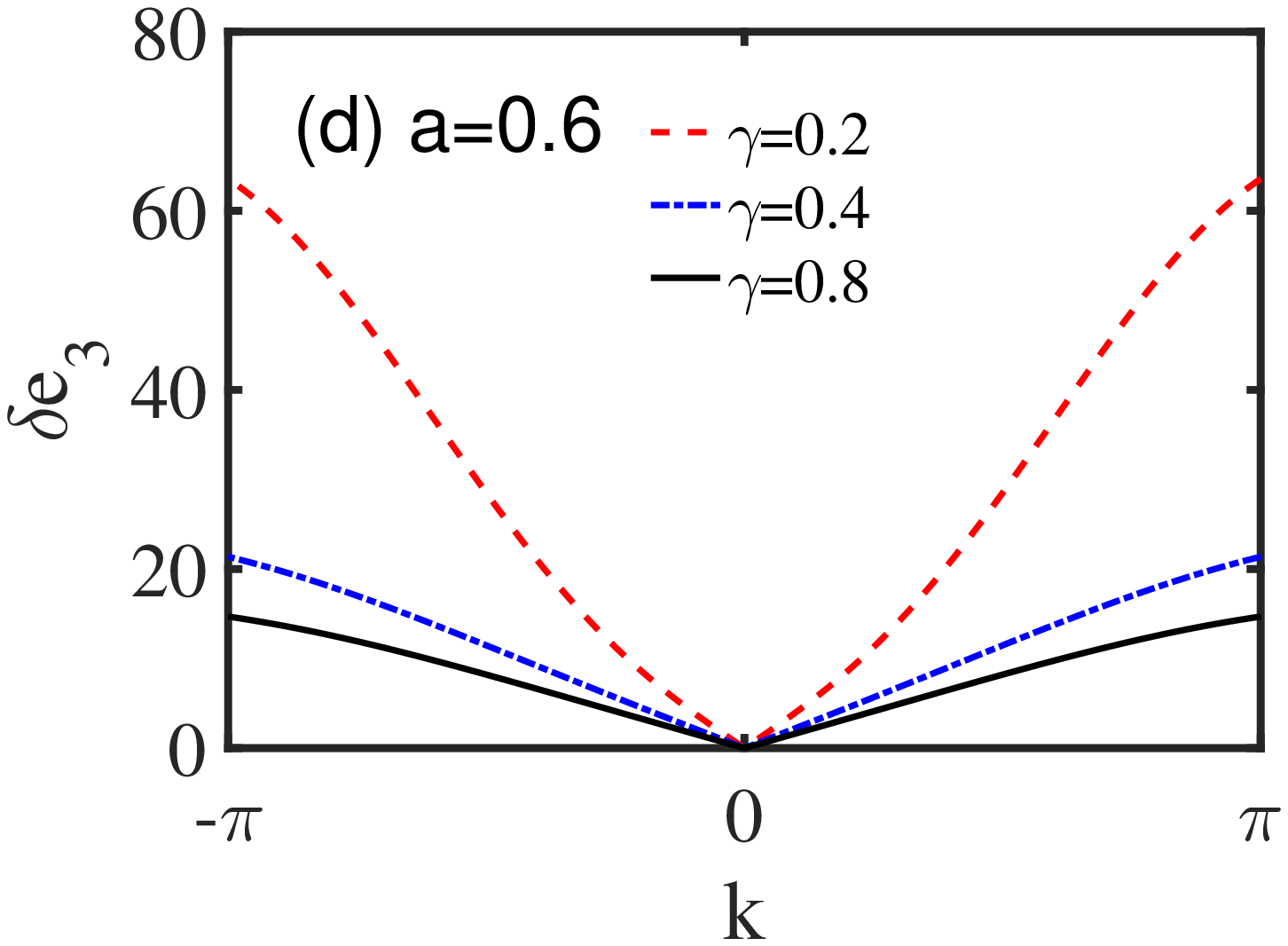}
\caption{(a) The distribution of $z$- and $w$-roots for the type III elementary excitation with $2N=8$, $n=3$, $a=0.2$ and $\gamma=0.6$.
(b) The excited energies (\ref{e03}) versus the different values of model parameter $a$ with $\lambda=0.2$ and $\gamma=0.2, 0.4, 0.8$.
(c) The type III excitation spectrum with $n=5$, $\gamma=0.6$ and $a=0, 0.6, 0.8$.
(d) The type III excitation spectrum with $n=5$, $a=0.6$ and $\gamma=0.2, 0.4, 0.8$. We shall note that the data in (b)-(d) are the results in the thermodynamic limit.}\label{fig-ee3}
\end{center}
\end{figure}

We now turn to the more general excitation determined by $2N-3$ real $z$-roots and one conjugate pair $\lambda \pm ni\gamma/2$ with $n\geq3$.
Accordingly, the related $w$-roots should be $2N-4$ real solutions and 2 conjugate pair with the form of $\lambda\pm(n+1)i\gamma/2$, $\lambda\pm(n-1)i\gamma/2$.
The roots distribution for $2N=8$ and $n=3$ is shown in Fig.\ref{fig-ee3}(a). Following the same procedure used above, we obtain the type III excitation energy as
\begin{eqnarray}
&&\delta e_3 = \frac{\cosh(4a)-\cos(2\gamma)}{\sin\gamma}
  \int
  \frac{\cos(2a\tau)\cos(2\lambda\tau)f(\tau)
  \tanh[(\pi-\gamma)\tau]} {\sinh(\pi\tau)}d\tau \nonumber \\ &&\qquad\;\;+
  \frac{\cosh(4a)-\cos(2\gamma)}{2\sin\gamma} \bigg[\frac{\sin(n-1)\gamma}
  {\cosh(2\lambda+2a)-\cos(n-1)\gamma}\nonumber\\
  &&\qquad\;\;+\frac{\sin(n-1)\gamma}{\cosh(2\lambda-2a)-\cos(n-1)\gamma}
  - \frac{\sin(n+1)\gamma}{\cosh(2\lambda+2a)-\cos(n+1)\gamma}\nonumber\\
  &&\qquad\;\;-\frac{\sin(n+1)\gamma}{\cosh(2\lambda-2a)-\cos(n+1)\gamma}\bigg], \label{e03}
\end{eqnarray}
where $f(\tau)=\cosh(y_{n-1}\tau)+\cosh(y_{n+1}\tau)$, $y_{n}=\pi-2\pi\delta_{n}$ and $\delta_n =n\gamma/(2\pi)-\lfloor n\gamma/(2\pi)\rfloor$ denoting the fractional part of $n\gamma/(2\pi)$.
The excitation energies with different values of model parameters $a$ and $\gamma$ are shown in Fig.\ref{fig-ee3}(b).
The momentum can be similarly calculated as
\begin{eqnarray}\label{k3}
&&\hspace{-1.7cm}k_3 = \frac{2i}{\pi-\gamma}\int\bigg\{ \frac{\cosh[x(z-\lambda)] \cos(xy_{n+1}/2)}{\cosh[2x(z-\lambda)] +\cos(xy_{n+1})}+\frac{\cosh[x(z-\lambda)] \cos(xy_{n-1}/2)}{\cosh[2x(z-\lambda)] +\cos(xy_{n-1})} \bigg \} \nonumber\\
    &&\hspace{-1cm} \times \ln\frac{\sinh(a+z-\frac{i\gamma}2)}{\sinh(a-z-\frac{i\gamma}2)}dz -i\ln\frac{\sinh(a+\lambda+\frac{n-1}2 i\gamma)\sinh(a+\lambda-\frac{n+1}2 i\gamma)}
    {\sinh(a-\lambda+\frac{n-1}2 i\gamma)\sinh(a-\lambda-\frac{n+1}2 i\gamma)} {~~}mod\,\{2\pi\},
\end{eqnarray}
where $x=\pi/(\pi-\gamma)$.
The dispersion relations with different values of model parameters $a$ and $\gamma$ are shown in Fig.\ref{fig-ee3}(c) and (d), respectively.
From them, we find that the nonlinearity of excitations becomes remarkable with the increasing of model parameters $a$ and $\gamma$.

\section{Conclusions}

In this paper, we have studied the exact solution of an integrable anisotropic spin chain with antiperiodic boundary condition,
where the interactions include the NN, NNN and chiral three-spin couplings.
We obtain the conserved topological momentum operator, energy spectrum and homogeneous BAEs.
In the thermodynamic limit, we calculate the ground state energy, three kinds of elementary excitations and dispersion relations
when the model parameter $a$ is real and anisotropic parameter $\eta$ is imaginary.
We find that due to the NNN and chirality interactions, the two peaks of excitation spectrum can locate away from the boundaries of Brillouin zone $\pm\pi$, which is quite different from that of the model with NN interaction.
Meanwhile, the nonlinearity of excitation spectrum can be enhanced.
We also note that the $t-W$ scheme and the new parametrization of transfer matrix provided in this paper can be generalized to study the model (\ref{Ham1}) with integrable non-diagonal boundary magnetic fields.

We shall note that when adding an external magnetic field along the $z$-direction, the system with periodic boundary condition is integrable.
Thus we can study the elementary excitations based on the corresponding exact solution with the similar method given in this paper.
Due to the existence of magnetic field, the spins would be polarized and the $Z_2$ invariance is broken.
Thus the patterns of $z$-roots are changed and the Fermi points are shifted.
Expanding the quasi-momentum at the points of new Fermi points,
we could obtain the nonlinear elementary excitations induced by the magnetic field. Other interesting physical quantities such as
dynamic structure factor and the spectral function can be calculated exactly \cite{Imam12}. The recent proposed pseudoparticle approach \cite{Car18} can also be used to
study both the static and the dynamical properties of the system.
For the antiperiodic boundary condition, the model Hamiltonian and the external magnetic field along the $z$-direction do not commutate with each other.
Thus the system is non-integrable. Due to the $U(1)$ symmetry broken, the eigenstates of the system are quite different from those with the periodic boundary condition.
The present eigenstates are the helical ones and the elementary excitations would show some interesting properties
such as the spinons are confined. Although the particle number with fixed spin state is not conserved, we can define
the topological conserved charge by combining the spin-up and spin-down states with the suitable coefficients. The helicity would be affected by the magnetic field. All these issues are worth studying.

\section*{Acknowledgments}

The financial supports from the National Key R\&D Program of China (Grants Nos. 2021YFA1400243 and 2016YFA0301500),
the National Natural Science Foundation of China (Grant Nos. 61835013, 11774150, 12074178, 12074410, 12047502, 11934015,
11975183, 11947301, 11774397 and 12147160), the Strategic
Priority Research Program of the Chinese Academy of Sciences (Grant Nos. XDB01020300, XDB21030300 and XDB33000000)
and the fellowship of China Postdoctoral Science Foundation (Grant No. 2020M680724) are gratefully acknowledged.

\end{document}